\documentclass[12pt]{article}
\usepackage{graphicx,amsmath}
\usepackage{bm}
\newcounter{mysubsubsection}

\newcommand{\spc}{{\ }}
\newcommand{\pr}[1]{{\sc{\lowercase{#1}}}}

\newcommand{\gras}[1]{\boldsymbol{#1}}

\newcommand{\codeversion}{2.49t}

\newcounter{leteq}

\newenvironment{eqnalpha}{\setcounter{leteq}{1}

\begin{eqnarray}}{\end{eqnarray}%
}

\newenvironment{eqnalphalabel}[1]{\setcounter{leteq}{1}
\raisebox{0cm}[0cm][0cm]{\begin{minipage}{1cm}%
\begin{eqnarray}\label{#1}&&\nonumber\end{eqnarray}\end{minipage}}

\begin{eqnarray}}{\end{eqnarray}%
}

\newcommand{\bnl}{\begin{eqnalpha}}
\newcommand{\enl}{\end{eqnalpha}}
\newcommand{\bnll}[1]{\begin{eqnalphalabel}{#1}}
\newcommand{\enll}{\end{eqnalphalabel}}

\newcommand{\keyw}{{\bf Keyword:}}
\newcommand{\key}[1]{\vspace{1ex}\noindent\keyw{\spc}{\tk{#1}}
                         \newline\phantom{\keyw{\spc}{\tk{#1}}}{\spc}}

\newcommand{\be}{\begin{equation}}
\newcommand{\ee}{\end{equation}}
\newcommand{\ba}{\begin{array}}
\newcommand{\ea}{\end{array}}
\newcommand{\bn}{\begin{eqnarray}}
\newcommand{\en}{\end{eqnarray}}
\newcommand{\bc}{\begin{center}}
\newcommand{\ec}{\end{center}}
\newcommand{\bi}{\begin{itemize}}
\newcommand{\ei}{\end{itemize}}

\newcommand{\tv}[1]{{\tt{#1}}\index{#1}}
\newcommand{\tk}[1]{{\tt{#1}}\index{#1}}
\newcommand{\ts}[1]{{\pr{#1}}\index{#1}}

\renewcommand{\tv}[1]{\textcolor{red}    {{\tt{#1}}}}
\renewcommand{\tk}[1]{\textcolor{blue}   {{\tt{#1}}}}
\renewcommand{\ts}[1]{\textcolor{green}  {{\pr{#1}}}}

\renewcommand{\tv}[1]{{\tt{#1}}}
\renewcommand{\tk}[1]{{\tt{#1}}}
\renewcommand{\ts}[1]{{\pr{#1}}}

\begin{document}

\vspace{0.5cm}
\begin{center}
        {\bf\Large
                     Solution of the Skyrme-Hartree-Fock-Bogolyubov equations in
                     the Cartesian deformed harmonic-oscillator basis. \\[1ex]
                     (VII) \pr{hfodd} (v\codeversion): a new version of the
program.
        }

\vspace{5mm}
        {\large
                       N. Schunck,$^{a,b,c}$\footnote
                       {E-mail: schunck1@llnl.gov}
                       J. Dobaczewski,$^{d,e}$
                       J. McDonnell,$^{b,c}$
                       W. Satu{\l}a,$^{d}$
                       J.A. Sheikh,$^{b,c}$
                       A. Staszczak,$^{b,c,f}$
                       M. Stoitsov,$^{b,c}$
                       P. Toivanen$^{e}$
        }

\vspace{3mm}
        {\it
          $^a$Physics Division, Lawrence Livermore National Laboratory
              Livermore, CA 94551, USA                                      \\
          $^b$Department of Physics and Astronomy, University of Tennessee,
              Knoxville, TN 37996, USA                                      \\
          $^c$Oak Ridge National Laboratory, P.O. Box 2008,
              Oak Ridge, TN 37831, USA                                      \\
          $^d$Institute of Theoretical Physics, Faculty of Physics, University of Warsaw,\\
                ul. Ho\.za 69, PL-00681 Warsaw, Poland                      \\
          $^e$Department of Physics, P.O. Box 35 (YFL),
              FI-40014 University of Jyv\"askyl\"a, Finland                 \\
          $^f$Department of Theoretical Physics, Maria Curie-Sk{\l}odowska University,\\
              pl. M. Curie-Sk{\l}odowskiej 1, 20-031 Lublin, Poland
        }
\end{center}

\vspace{5mm}
\hrule

\vspace{2mm}
\noindent{\bf Abstract}

We describe the new version (v\codeversion) of the code \pr{hfodd} which solves
the nuclear Skyrme Hartree-Fock (HF) or Skyrme Hartree-Fock-Bogolyubov (HFB)
problem by using the Cartesian deformed harmonic-oscillator basis. In the new
version, we have implemented the following physics features:
(i)    the isospin mixing and projection,
(ii)   the finite temperature formalism for the HFB and HF+BCS methods,
(iii)  the Lipkin translational energy correction method,
(iv)   the calculation of the shell correction.
A number of specific numerical methods have also been implemented in order to
deal with large-scale multi-constraint calculations and hardware limitations:
(i)    the two-basis method for the HFB method,
(ii)   the Augmented Lagrangian Method (ALM) for multi-constraint calculations,
(iii)  the linear constraint method based on the approximation of the RPA matrix
for multi-constraint calculations,
(iv)   an interface with the axial and parity-conserving Skyrme-HFB code \pr{hfbtho},
(v)    the mixing of the HF or HFB matrix elements instead of the HF fields.
Special care has been paid to using the code on massively parallel
leadership class computers. For this purpose, the following features are now
available with this version:
(i)    the Message Passing Interface (MPI) framework
(ii)   scalable input data routines
(iii)  multi-threading via OpenMP pragmas
(iv)   parallel diagonalization of the HFB matrix in the simplex breaking case
using the ScaLAPACK library.
Finally, several little significant errors of the previous published version were
corrected.
\vspace{2mm}
\hrule

\vspace{2mm}

\vspace{5mm}
{\bf\large NEW VERSION PROGRAM SUMMARY}

\bigskip\noindent{\it Title of the program:} \pr{hfodd}
                 (v\codeversion)

\bigskip\noindent{\it Catalogue number:}
                   ....

\bigskip\noindent{\it Program obtainable from:}
                      CPC Program Library, \
                      Queen's University of Belfast, N. Ireland
                      (see application form in this issue)

\bigskip\noindent{\it Reference in CPC for earlier version of program:}
                      J. Dobaczewski, W. Satu{\l}a, B.G. Carlsson, J. Engel, P.
Olbratowski,
                      P. Powa{\l}owski, M. Sadziak, J. Sarich, N. Schunck, A.
Staszczak,
                      M. Stoitsov, M. Zalewski, H. Zdu{\'n}czuk,
                      Comput.\ Phys.\ Commun.\ {\bf 180} (2009) 2361 (v2.40h).

\bigskip\noindent{\it Catalogue number of previous version:}
                  ADFL\_v2\_1

\bigskip\noindent{\it Licensing provisions:} GPL v3

\bigskip\noindent{\it Does the new version supersede the previous one:} yes

\bigskip\noindent{\it Computers on which the program has been tested:}
                      Intel Pentium-III, Intel Xeon, AMD-Athlon, AMD-Opteron,
Cray XT4, Cray XT5

\bigskip\noindent{\it Operating systems:} UNIX, LINUX, Windows$^{\text{xp}}$

\bigskip\noindent{\it Programming language used:} FORTRAN-90

\bigskip\noindent{\it Memory required to execute with typical data:} 10 Mwords

\bigskip\noindent{\it No. of bits in a word:}
The code is written in single-precision for the use on a 64-bit processor. The
compiler option {\tt{}-r8} or {\tt{}+autodblpad} (or equivalent) has to be used
to promote all real and complex single-precision floating-point items
to double precision when the code is used on a 32-bit machine.

\bigskip\noindent{\it Has the code been vectorised?:} Yes

\bigskip\noindent{\it Has the code been parallelized?:} Yes

\bigskip\noindent{\it No.{\spc}of lines in distributed program:}
                      104 666 (of which 47 059 are comments and separators)

\bigskip\noindent{\it Keywords:}
                      Hartree-Fock; Hartree-Fock-Bogolyubov; Skyrme interaction;
                      Self-consistent mean field;
                      Nuclear many-body problem; Superdeformation;
                      Quadrupole deformation; Octupole deformation; Pairing;
                      Nuclear radii; Single-particle spectra;
                      Nuclear rotation; High-spin states;
                      Moments of inertia; Level crossings; Harmonic oscillator;
                      Coulomb field; Pairing; Point symmetries;
                      Yukawa interaction; Angular-momentum projection;
                      Generator Coordinate Method; Schiff moments;
                      Isospin mixing; Isospin projection, Finite temperature;
                      Shell correction; Lipkin method; Multi-threading; Hybrid
                      programming model; High-performance computing.

\bigskip\noindent{\it Nature of physical problem}

\noindent
The nuclear mean field and an analysis of its symmetries in realistic cases are
the main ingredients of a description of nuclear states. Within the Local
Density Approximation, or for a zero-range velocity-dependent Skyrme
interaction, the nuclear mean field is local and velocity dependent. The
locality allows for an effective and fast solution of the self-consistent
Hartree-Fock equations, even for heavy nuclei, and for various nucleonic
($n$-particle $n$-hole) configurations, deformations, excitation energies, or
angular momenta. Similarly, Local Density Approximation in the particle-particle
channel, which is equivalent to using a zero-range interaction, allows for a
simple implementation of pairing effects within the Hartree-Fock-Bogolyubov
method.

\bigskip\noindent{\it Method of solution}

\noindent
The program uses the Cartesian harmonic oscillator basis to expand
single-particle or single-quasiparticle wave functions of neutrons and protons
interacting by means of the Skyrme effective interaction and zero-range pairing
interaction. The expansion coefficients are determined by the iterative
diagonalization of the mean-field Hamiltonians or Routhians which depend
non-linearly on the local neutron and proton densities.  Suitable constraints
are used to obtain states corresponding to a given configuration, deformation or
angular momentum. The method of solution has been presented in: J. Dobaczewski
and J. Dudek, Comput.\ Phys.\ Commun.\ {\bf 102} (1997) 166.

\bigskip\noindent{\it Summary of revisions}

\noindent
\begin{enumerate}
\setlength{\itemsep}{-1ex}
\item Isospin mixing and projection of the HF states has been implemented.
\item The finite-temperature formalism for the HFB equations has been
implemented.
\item The Lipkin translational energy correction method has been implemented.
\item Calculation of the shell correction has been implemented.
\item The two-basis method for the solution to the HFB equations has been
implemented.
\item The Augmented Lagrangian Method (ALM) for calculations with multiple constraints
has been implemented.
\item The linear constraint method based on the cranking approximation of the
RPA matrix has been implemented.
\item An interface between \pr{hfodd} and the axially-symmetric and parity-conserving
code \pr{hfbtho} has been implemented.
\item The mixing of the matrix elements of the HF or HFB matrix has been
implemented.
\item A parallel interface using the MPI library has been implemented.
\item A scalable model for reading input data has been implemented.
\item OpenMP pragmas have been implemented in three subroutines.
\item The diagonalization of the HFB matrix in the simplex-breaking case has
been parallelized using the ScaLAPACK library.
\item Several little significant errors of the previous published version were
corrected.
\end{enumerate}

\bigskip\noindent{\it Restrictions on the complexity of the problem}

\noindent

\bigskip\noindent{\it Typical running time}

\noindent

\bigskip\noindent{\it Unusual features of the program}

\noindent
The user must have access to (i) the NAGLIB subroutine \pr{F02AXE}, or LAPACK
subroutines \pr{ZHPEV}, \pr{ZHPEVX}, \pr{ZHEEVR}, or \pr{ZHEEVD}, which
diagonalize complex hermitian matrices, (ii) the LAPACK subroutines \pr{DGETRI}
and \pr{DGETRF} which invert arbitrary real matrices, (iii) the LAPACK subroutines
\pr{DSYEVD}, \pr{DSYTRF} and \pr{DSYTRI} which compute eigenvalues and
eigenfunctions of real symmetric matrices and (iv) the LINPACK subroutines
\pr{ZGEDI} and \pr{ZGECO}, which invert arbitrary complex matrices and calculate
determinants, (v) the BLAS routines \pr{DCOPY}, \pr{DSCAL}, \pr{DGEEM} and
\pr{DGEMV} for double-precision linear algebra and \pr{ZCOPY}, \pr{ZDSCAL},
\pr{ZGEEM} and \pr{ZGEMV} for complex linear algebra, or provide another set
of subroutines that can perform such tasks. The BLAS and LAPACK subroutines
can be obtained from the Netlib Repository at the University of Tennessee,
Knoxville: \verb+http://netlib2.cs.utk.edu/+.

\bigskip

\bigskip

{\bf\large LONG WRITE-UP}

\bigskip


\section{Introduction}
\label{sec:intro}

The method of solving the Hartree-Fock (HF) equations in the Cartesian harmonic
oscillator (HO) basis was described in the publication, Ref.~\cite{[Dob97c]}.
Five versions of the code \pr{hfodd} were previously published:
(v1.60r)~\cite{[Dob97d]},(v1.75r)~\cite{[Dob00c]}, (v2.08i)~\cite{[Dob04]},
(v2.08k)~\cite{[Dob05]}, and (v2.40h)~\cite{[Dob09d]}. The User's Guide for
version (v2.40v) is available in Ref.~\cite{[Dob09]} and the code home page is
at \verb+http://www.fuw.edu.pl/~dobaczew/hfodd/hfodd.html+. The present paper is
a long write-up of the new version (v\codeversion) of the code \pr{hfodd}. This
extended version features the isospin mixing and projection of the HF states,
the finite-temperature formalism for the HF+BCS and HFB equations,
and several other major modifications. It is also built upon a hybrid MPI/OpenMP
parallel programing model which allows large-scale calculations on massively
parallel computers. In serial mode, it remains fully compatible with all
previous versions. Information provided in previous publications
\cite{[Dob97d]}-\cite{[Dob09d]} thus remains valid, unless explicitly mentioned
in the present long write-up.

In Section \ref{sec:modifs} we briefly review the modifications introduced in
version (v\codeversion) of the code \pr{hfodd}. We distinguish between features
implementing (i) new physics modeling capabilities, (ii) new numerical
techniques and (iii) parallel computing methods. Section \ref{sec:input} lists
all additional new input keywords and data values, introduced in version
(v\codeversion). In serial mode, the structure of the input data file remains
the same as in the previous versions, see Section 3 of Ref.~\cite{[Dob97d]}. In
parallel mode, two input files, with strictly enforced names, must be used: {\tt
hfodd.d} has the same keyword structure as all previous \pr{hfodd} input files,
with the restriction that not all keywords can be activated (see list in
Sec.~\ref{subsubsec:hfodd.d}); {\tt hfodd\_mpiio.d} contains processor-dependent
data, see Sec.~\ref{subsubsec:hfodd_mpiio.d}.


\section{Modifications introduced in version (v\codeversion)}
\label{sec:modifs}


\setcounter{mysubsubsection}{0}

\subsection{New Physics Features}
\label{subsec:physics}


\subsubsection{Isospin Mixing and Projection}
\label{subsubsec:isospin}

The concept of isospin symmetry, having its roots in the approximate charge
independence of the nucleon-nucleon interaction, was already introduced in nuclear
physics in the $1930$s by Heisenberg and
Wigner~\cite{[Hei32a],[Wig37]}. Throughout the years, it has proven to be
extremely powerful and not abated by the presence of the Coulomb force -- the
main source of the isospin symmetry violation in nuclei -- simply because the
isovector and isotensor parts of the Coulomb force are much weaker than the
dominant, isospin symmetry preserving components of the Coulomb and strong
interactions.

Apart from the explicit violation of the isospin symmetry due to the strong and,
predominantly, Coulomb interactions, various approximate theoretical methods
used in nuclear structure calculations are often the sources of unphysical
violation of this symmetry by
themselves~\cite{[Eng70],[Cau80],[Raf09c],[Sat09a],[Sat10]}. This specifically
concerns the Hartree-Fock and Kohn-Sham theories that employ
independent-particle wave functions, which manifestly break the isospin symmetry
in $N\neq Z$ nuclei even for isospin-conserving interactions. The most prominent
effects of this spontaneous isospin-symmetry-breaking occur, however, in the
ground-state configuration of odd-odd $N=Z$ nuclei and in $T\neq0$ excited
configurations of $N=Z$ nuclei, see Ref.~\cite{[Sat10]} and references cited
therein. Hence, practical implementation of the method requires the
isospin projection and subsequent rediagonalization of the entire
Hamiltonian in the isospin-projected basis. These two major
building blocks of the isospin projection method will be described below.
The discussion will be followed by a short presentation of the
extended version of our model including the isospin and angular-momentum
projections which is needed for specific applications including
calculation of the isospin-symmetry breaking corrections to superallowed
$\beta-$decay~\cite{[Sat10a],[Sat10b],[Sat11a]}.

{\bf The isospin projection: } To remove the unphysical isospin-symmetry
violation introduced by the mean-field (MF) approximation, the code {\sc
hfodd} (v\codeversion) was equipped with a new tool allowing for the isospin
projection after variation of an arbitrary symmetry-unrestricted Slater
determinant $| \Phi \rangle$ provided by the code. The method implemented 
uses the standard one-dimensional isospin-projection operator
$\hat{P}^T_{T_z T_z}$:
\begin{eqnarray}\label{eqn:imk}
|TT_z\rangle & = &  \frac{1}{\sqrt{N_{TT_z}}}\hat{P}^T_{T_z T_z} |\Phi\rangle
= \frac{2T+1}{2 \sqrt{N_{TT_z}}}\int_0^\pi d\beta_T\; \sin\beta_T \; d^{T}_{T_z
T_z}(\beta_T )\;
\hat{R}(\beta_T )|\Phi\rangle ,
\end{eqnarray}
which allows for decomposing the Slater determinant $| \Phi \rangle$,
\begin{equation}\label{mix}
|\Phi \rangle = \sum_{T\geq |T_z|}b_{TT_z}|TT_z\rangle,
\end{equation}
into good-isospin basis $|TT_z\rangle$. Here,  $\beta_T$ denotes the Euler angle
associated with the rotation operator $\hat{R}(\beta_T )= e^{-i\beta_T
\hat{T}_y}$ about the $y$-axis in the isospace, $d^{T}_{T_z T_z}(\beta_T )$ is
the Wigner function~\cite{[Var88]}, and $T_z =(N-Z)/2$ is the third component of
the total isospin $T$. The  normalization factors $N_{TT_z}$, or interchangeably
the expansion coefficients $b_{TT_z}$, read:
\begin{eqnarray}
\label{eqn:ovr}
N_{TT_z}  & \equiv &  |b_{TT_z}|^2 = \langle \Phi | \hat{P}^T_{T_z T_z} | \Phi
\rangle
= \frac{2T+1}{2}\int_0^\pi d\beta_T \sin\beta \; d^{T}_{T_z T_z}
(\beta_T ) \; {\mathcal N}(\beta_T),
\end{eqnarray}
where
\begin{equation}
{\mathcal N}(\beta_T)  = \langle \Phi| \hat{R}(\beta_T)| \Phi\rangle
\label{N-ker}
\end{equation}
stands for the overlap kernel.

The isospin projection operator is used to construct a subspace (basis) of
good-isospin states $|TT_z\rangle$. Its size is controlled by the parameter
$\varepsilon_T$, such that only the states $|TT_z\rangle$ that have tangible
contributions, $|b_{TT_z}|^2 \geq \varepsilon_T$, to the MF state are retained
for further rediagonalization. In practice, $\varepsilon_T=10^{-10}$ sets the
limit of $T\leq |T_z|+5$. The good-isospin basis created in this way, although
of rather small dimension, is believed to capture the right balance between the
short-range strong interaction and the long-range Coulomb force (see discussion
in Ref.~\cite{[Sat09a]}).  The parameter $\varepsilon_T = 10^{-10}$ also ensures
that the two basic quantities reflecting the accuracy of the method, namely, the
overlap (normalization) sum rule,
\begin{equation}\label{mix2}
\sum_{T\geq |T_z|} |b_{TT_z}|^2 = 1,
\end{equation}
and total MF energy sum rule,
\begin{equation}\label{e-sumrule}
E_{MF} \equiv \langle \Phi |\hat H | \Phi \rangle
 =  \sum_{TT^\prime \geq |T_z|}  b_{T^\prime T_z}^{*} b_{TT_z}
\langle T^\prime T_z|\hat H | TT_z\rangle ,
\end{equation}
are both fulfilled with extremely high accuracy. The latter property is due to
the fact that the isospin-projection method is practically free from divergences
plaguing particle-number and angular-momentum~\cite{[Ang01],[Rob07],[Dob07d],
[Zdu07]} methods. An analytical proof of this rather remarkable feature of the
isospin projection is given in Ref.~\cite{[Sat10]}.

{\bf Rediagonalization in the isospin projected basis:} Expansion coefficients
$b_{T,T_z}$ do not reflect the physical isospin mixing. Indeed, they are
affected by the spurious isospin mixing, which is due to the spontaneous
breaking of the isospin symmetry caused by the MF approximation. To calculate
the true isospin mixing, one needs to rediagonalize the total nuclear
Hamiltonian in the good-isospin basis $|TT_z\rangle$. The present implementation
of the code \pr{hfodd} admits Hamiltonians that include the isoscalar part of
the kinetic energy $\hat T$, isospin-invariant Skyrme functional, $\hat V^{S}$,
and Coulomb force, $\hat V^C$; the latter can further be decomposed into the
isoscalar, $\hat V_{00}^C$, isovector, $\hat V_{10}^C$, and isotensor, $\hat
V_{20}^C$, components, that is,
\begin{eqnarray}\label{ham} \hat H & =
&\hat T + \hat V^{S} + \hat V^{C} \equiv  \hat T + \hat V^{S}
+\hat V_{00}^C +  \hat V_{10}^C + \hat V_{20}^C ,
\end{eqnarray}
where
\begin{eqnarray}
     \hat  V_{00}^C(r_{ij}) & = & \; \; \; \frac{1}{4}\frac{e^2}{r_{ij}}
       \left( 1 + \frac{1}{3} \hat {\tau}^{(i)} \circ
                  \hat {\tau}^{(j)} \right) , \label{C00} \\
      \hat V_{10}^C(r_{ij}) & = &  - \frac{1}{4}\frac{e^2}{r_{ij}}
          \left( \hat \tau_{10}^{(i)}  +
                  \hat \tau_{10}^{(j)} \right) , \label{C10}   \\
     \hat  V_{20}^C(r_{ij}) & = &  \; \; \; \frac{1}{4}\frac{e^2}{r_{ij}} \;
              \left( \hat \tau_{10}^{(i)}
         \hat \tau_{10}^{(j)} - \frac{1}{3} \hat {{\tau}}^{(i)} \circ
                                            \hat {{\tau}}^{(j)} \right) .
                                \label{C20}
\end{eqnarray}
Note, that the components $\hat V_{\lambda 0}^C$ are constructed by coupling the
spherical components of the one-body isospin operator:
\begin{equation}
      \hat  \tau_{10} = \hat \tau_{z}, \quad  \hat \tau_{1\pm 1}
           = \mp \frac{1}{\sqrt{2}}
        \left( \hat \tau_{x} \pm i \hat \tau_{y} \right) ,
\end{equation}
where $\hat \tau_{i},\,i=x,y,z$ denote Pauli matrices and symbol $\circ$ stands
for the scalar product of isovectors. Hence, from a mathematical viewpoint, they
represent isoscalar, covariant rank-1 (isovector), and covariant rank-2 axial
(isotensor) spherical tensor components of the Coulomb interaction,
respectively. This mathematical property allows the use of Racah algebra in order to
calculate matrix elements of the Hamiltonian. The rather lengthy details
concerning this specific theoretical aspect of our model are given in
Ref.~\cite{[Sat10]} and will not be repeated here.

Rediagonalization of the total Hamiltonian in the good-isospin basis leads to
the eigenstates:
\begin{equation}\label{mix2a}
|n,T_z\rangle
= \sum_{T\geq |T_z|}a^n_{TT_z}|TT_z\rangle ,
\end{equation}
numbered by index $n$. Apart of the eigenenergies $E_{n,T_z}$ and the amplitudes
$a^n_{TT_z}$ that define the degree of isospin mixing, the code also provides
the so-called isospin (Coulomb) mixing coefficients or, equivalently, the
isospin impurities. For the $n$-th eigenstate, the isospin impurity is defined
as $\alpha_C^n = 1 - |a^n_{TT_z}|_{\text{max}}^2$, where
$|a^n_{TT_z}|_{\text{max}}^2$ stands for the squared norm of the dominant
amplitude in the wave function $|n,T_z\rangle$, and is given in percents.

Evaluation of the isospin impurity $\alpha_C$ is a prerequisite for determining
the isospin-breaking corrections $\delta_C$ to the $0^+  \rightarrow 0^+$ Fermi
matrix element of the isospin raising/lowering operator $\hat T_{\pm}$. Of
particular interest in nuclear physics are the Fermi matrix elements:
\begin{equation}\label{fermi}
|\langle I=0, T\approx 1, T_z = \pm 1 | \hat T_{\pm} | I=0, T\approx 1, T_z = 0
\rangle |^2
\equiv 2 ( 1-\delta_C ) ,
\end{equation}
for a set of nuclei undergoing the super-allowed beta decay, because the
$\delta_C$ parameter is the key nuclear quantity needed for precise nuclear
tests of the conserved-vector-current hypothesis and for the determination of
the up-down matrix element in the Cabibbo-Kobayashi-Maskawa matrix (see
Ref.~\cite{[Tow08]} and references quoted therein). The calculation of the Fermi
matrix elements (\ref{fermi}) was one of the primary motivations to couple the
newly developed isospin projection with the existing angular-momentum
projection~\cite{[Dob09d]}. Indeed, such a four-dimensional projection appears
to be absolutely necessary to get reliable representation of decaying states in
daughter (parent) $|I=0, T\approx 1, T_z = \pm 1 \rangle$ and parent (daughter)
$| I=0, T\approx 1, T_z = 0 \rangle$ nuclei undergoing the super-allowed
transition, respectively (see Ref.~\cite{[Sat10a]}). It should be stressed,
however, that the range of applicability of the four-dimensional projection is
by no means limited to the computation of matrix elements (\ref{fermi}) but also
encompasses, in particular, various applications in high-spin physics in
$N\sim Z$ nuclei.

{\bf The four-dimensional isospin and angular momentum projection:}
The implementation of the four-dimensional projection follows rather closely the
angular-momentum projection scheme adopted in version (v2.40h) of the code and
described in detail in Ref.~\cite{[Dob09d]} (cf.
Refs.~\cite{[Zdu07],[Zdu07a]}. Hence, in the following, we will refrain from
technicalities and concentrate on discussing the main building blocks of the
method. The starting point is the good angular momentum $I$ and good isospin
$T$ basis generated by acting with standard angular-momentum $\hat P^I_{MK}$
and isospin $\hat P^T_{T_z T_z}$ projectors on the Slater determinant $|\Phi
\rangle$: \begin{equation}\label{ITbasis} |IMK; TT_z\rangle = \hat P^T_{T_z
T_z} \hat P^I_{MK} |\Phi \rangle , \end{equation} where $M$ and $K$ stand for
the angular-momentum projections along the laboratory and intrinsic $z$-axes,
respectively. The basis composed of states $|IMK; TT_z\rangle$ is
over-complete. This problem is overcome by constructing, separately for each
$I$ and $T$, the so-called collective subspace spanned by the natural states:
\begin{equation}
\label{nat_st}   |IM; TT_z \rangle^{(m)} =
  \frac{1}{\sqrt{n_m}} \sum_{K} \eta_K^{(m)}
  |IMK; TT_z\rangle.
\end{equation}
The $m^{\text{th}}$ natural state is constructed by using the mixing amplitudes
$\eta_{K}^{(m)}$ that correspond to the $m^{\text{th}}$ eigenstate of the norm
matrix, see Eq. (9) in \cite{[Dob09d]}:
\be\label{egn:eignorm}
\sum_{K'} N^{TT_{z}}_{K K'} {\eta}^{(m)}_{K'} = n_m\; {\eta}^{(m)}_{K}.
\ee
Only the eigenstates corresponding to eigenvalues $n_m > \zeta$ are taken into
account, with the basis cut-off parameter $\zeta$ introduced in
Ref.~\cite{[Dob09d]}. In this way, for each value of the angular momentum $I$
and isospin $T$, the collective subspace contains $m_{\text{max}}(I,T)$ states.
The overlap matrix appearing in Eq.~(\ref{egn:eignorm}) reads:
\begin{eqnarray}\label{egn:eignorm2}
N^{TT_{z}}_{K K'} & = &\langle \Phi | \hat P^T_{T_z T_z}  \hat P^I_{K K^\prime}
| \Phi
\rangle \nonumber = \\ &=&
\frac{(2I+1)(2T+1)}{16\pi^2}\int d\beta_T\;
d^{T}_{T_z T_z} (\beta_T ) \int d\Omega\; D^{I*}_{KK'} (\Omega ) \; \langle
\Phi| \hat{R}(\beta_T) \hat{R}(\Omega)| \Phi\rangle ,
\end{eqnarray}
where $\hat{R}(\Omega) = e^{-i\alpha \hat{I}_z}e^{-i\beta \hat{I}_y} e^{-i\gamma
\hat{I}_z}$  stands for the space-rotation operator, which depends on three
Euler angles $\Omega = (\alpha, \beta, \gamma)$, and $D^{I}_{KK'} (\Omega )$ is
the Wigner function.

To simultaneously take into account the $K$-mixing and isospin mixing, the code
performs, separately for each value of the angular momentum $I$, the full
diagonalization of the total Hamiltonian (\ref{ham}) in the $n(I)$-dimensional,
$n(I) =\sum_{T \geq |T_z|} m_{\text{max}}(I,T)$, collective space spanned by
natural states (\ref{nat_st}). Such a diagonalization leads to the eigenstates
of the form:
\begin{equation}   \label{KTmix} |n; IM; T_z\rangle =
\sum_{T\geq |T_z|} \sum_{m=1}^{m_{\text{max}}(I,T)}
   a^{(n)}_{mT}(I) |IM; TT_z\rangle^{(m)} ,
\end{equation}
which are labeled by the conserved quantum numbers $I,M$, and $T_z=(N-Z)/2$, and
by the additional index $n$, which characterizes the $K$ and isospin mixing. For
the sake of completeness, it is worth mentioning that the code also provides the
$K$-mixed and isospin-conserving eigenstates, which result from the
diagonalization of the total Hamiltonian with all isospin-symmetry-breaking
($\Delta T \ne 0$) matrix elements set to zero.


\subsubsection{Finite-temperature Formalism}

The equilibrium state of a physical system at constant temperature $T$ and
chemical potential $\lambda$ is obtained from the minimization of the grand
canonical potential $\Omega$ \cite{[Goo81],[Die81],[Bon84],[Egi00],[Mar03]}
\begin{equation}
\Omega  = E - TS - \lambda N  ,  \label{grpot}
\end{equation}
where the energy ($E$), entropy ($S$) and particle-number ($N$) are statistical
averages and are given by
\begin{eqnarray}
E & = &    {\rm Tr} (\mathcal{\hat D} \hat H)  ,                    \label{ene}
\\
S & = & -k {\rm Tr} (\mathcal{\hat D} {\rm ln} \mathcal{\hat D} ) , \label{entr}
\\
N & = &    {\rm Tr} (\mathcal{\hat D} \hat N)  .
\label{partn}
\end{eqnarray}
The density operator $\mathcal{\hat D}$ and the grand partition function $\mathcal{Z}$ are defined, respectively,
as
\begin{eqnarray}
\mathcal{\hat D} & = & \displaystyle\frac{1}{\mathcal{Z}} e^{-\beta (\hat H -
\lambda \hat N)} \,\,\, , \label{denmat} \\
\mathcal{Z} & = & {\rm Tr}\left[ e^{-\beta (\hat H - \lambda \hat N)} \right]
\,\,\, ,
\end{eqnarray}
where $\beta = 1/kT$ and $\hat{H}$ is the two-body Hamiltonian. In the
MF approximation, the two-body density operator in Eq. (\ref{denmat}) is
replaced by a one-body operator. It has been demonstrated in \cite{[Goo81]} that
the variation of the grand canonical potential with respect to the density
operator $\hat{\mathcal{D}}$ leads to HFB equations that are formally equivalent
to the $T=0$ equations, namely
\begin{equation}
\left( \begin{array}{cc}
h-\lambda       &           \Delta \\
 -\Delta^{\ast} & -h^{\ast}+\lambda
\end{array}\right)
\left( \begin{array}{c} U_{\mu}  \\
                        V_{\mu}
       \end{array}\right) = E_{\mu}
\left(
       \begin{array}{c} U_{\mu} \\
                        V_{\mu}
       \end{array}\right),
\label{hfbeq}
\end{equation}
where $h$ and $\Delta$ are the Hartree-Fock and pairing potentials, and are
obtained from the energy density functional as usual. The inclusion of finite
temperature in the formalism is achieved by generalizing the expression of the
density matrix and pairing tensor. In configuration space, they
read~\cite{[Mar03]}
\begin{eqnarray}
\rho   & = & U f U^{\dagger} + V^{\ast} (1-f) V^{T}, \\
\kappa & = & U f V^{\dagger} + V^{\ast} (1-f) U^{T},
\label{densities_matrix}
\end{eqnarray}
where the quantity ``$f$'' stands for the Fermi function, defined as,
\begin{equation}
f_{\mu} = \frac{1}{1+e^{\beta E_\mu}} \,\,\, , \label{fermi-dirac}
\end{equation}
with $E_{\mu}$ the $\mu^{\text{th}}$ quasi-particle energy. In coordinate space,
their expression becomes
\begin{eqnarray}
\label{densities_space}
\rho(\gras{r}\sigma,\gras{r'}\sigma')   & = &
\sum_{0 \leq E_{\mu} \leq E_{\text{max}}}
\left\{
f_{\mu} U^{(\mu)}(\gras{r}\sigma) U^{(\mu)*}(\gras{r'}\sigma')
+
(1 - f_{\mu})V^{(\mu)*}(\gras{r}\sigma) V^{(\mu)}(\gras{r'}\sigma')
\right\}, \\
\kappa(\gras{r}\sigma,\gras{r'}\sigma') & = &
\sum_{0 \leq E_{\mu} \leq E_{\text{max}}}
\left\{
f_{\mu} U^{(\mu)}(\gras{r}\sigma) V^{(\mu)*}(\gras{r'}\sigma')
+
(1 - f_{\mu}) V^{(\mu)*}(\gras{r}\sigma) U^{(\mu)}(\gras{r'}\sigma')
\right\}.
\end{eqnarray}
All additional densities (kinetic, spin-current, etc.) are derived from the
particle density (\ref{densities_space}). In \pr{hfodd}, the conventional
pairing tensor is replaced by the pairing density
$\tilde{\rho}(\gras{r}\sigma,\gras{r'}\sigma')$ obtained according to:
$\tilde{\rho}(\gras{r}\sigma,\gras{r'}\sigma') = -2\sigma'
\kappa(\gras{r}\sigma,\gras{r'} -\sigma') $, see \cite{[Dob84]}.

The code \pr{hfodd} implements the finite-temperature formalism in the HFB and
HF+BCS modes. For the latter case, we refer to Sec. 5 of \cite{[Goo81]} for the
details of the expressions coded. For the former case, we would like to
emphasize that the calculation of the Fermi level $\lambda$ needs to
be modified at $T>0$. Let us
recall that the adjustment of $\lambda$ in \pr{hfodd} is based on BCS formula
\cite{[Dob84]}: given the equivalent spectrum of single-particle (s.p.) states
$\varepsilon_{\mu}$ and pairing gaps $\Delta_{\mu}$, the particle number is
computed according to:
\begin{equation}
N = \sum_{\mu} \left[ v_{\mu}^{2} + (u_{\mu}^{2} - v_{\mu}^{2}) f_{\mu} \right],
\end{equation}
with the Fermi functions $f_{\mu}$ of Eq. (\ref{fermi-dirac}) and the occupation
factors given by:
\begin{equation}
v_{\mu}^{2} = \frac{1}{2} \left[ 1 - \frac{\varepsilon_{\mu} -
\lambda}{E^{\text{BCS}}_{\mu}} \right],\ \ \ \
u_{\mu}^{2} = 1 - v_{\mu}^{2},
\end{equation}
with $E^{\text{BCS}}_{\mu} = \sqrt{(\varepsilon_{\mu} - \lambda)^{2} +
\Delta_{\mu}^{2}}$. When applying the Newton-Raphson method to determine
$\lambda$ by the condition that $N = N_{0}$, the implicit dependence of the
$f_{\mu}$ on $\lambda$ must be taken into account. This is done by updating the
$f_{\mu}$ at each $\lambda$ according to:
\begin{equation}
f_{\mu}(\lambda) = \frac{1}{1 + e^{\beta E^{\text{BCS}}_{\mu} }},
\end{equation}
and introducing the corresponding additional term $\partial
f_{\mu}/\partial\lambda$ in the derivative $\partial N/ \partial\lambda$. The
contribution from the thermal occupation factors is crucial for the convergence in
the unpaired regime. Note that in the case of zero-range pairing interactions, there
can be stability issues for low cut-offs near the phase transition. Recently, the finite
temperature extension of \pr{hfodd} has been used in a systematic study of fission
paths and barriers of actinide and superheavy elements \cite{[Pei09],[She09]}.


\subsubsection{Lipkin Translational Energy Correction}

According to the Lipkin method \cite{[Lip60],[Dob09g]}, the linear-momentum
projected energy of a system at rest can be calculated without the actual
projection as:
\begin{equation}
E_{\Phi}(0)= \langle \Phi \mid \mathrm{\hat{H}}-\mathrm{\hat{K}} \mid \Phi
\rangle, \label{Projected Energy}
\end{equation}
where $\mathrm{\hat{H}}$ is the two-body effective Hamiltonian and
\begin{equation}
\mathrm{\hat{K}}=k\mathbf{\hat{P}}^2
\label{eq:lipkin}
\end{equation}
is the Lipkin operator in the quadratic approximation, $\mathbf{\hat{P}} =
\sum_{i=1}^{A} \mathbf{\hat{p}}_{i}$ is the total linear momentum operator and
$k$ is a parameter to be determined. The optimum state is found by minimizing
the right hand side of Eq.~(\ref{Projected Energy}). Note that the projected
energy depends parametrically on the parameter $k$, that is, there is no variation
with respect to $k$ (no contribution to the HF potentials).

\begin{figure}[h]
\begin{center}
\begin{minipage}[t]{0.50\textwidth}
\begin{center}
\includegraphics[width=0.95\textwidth]{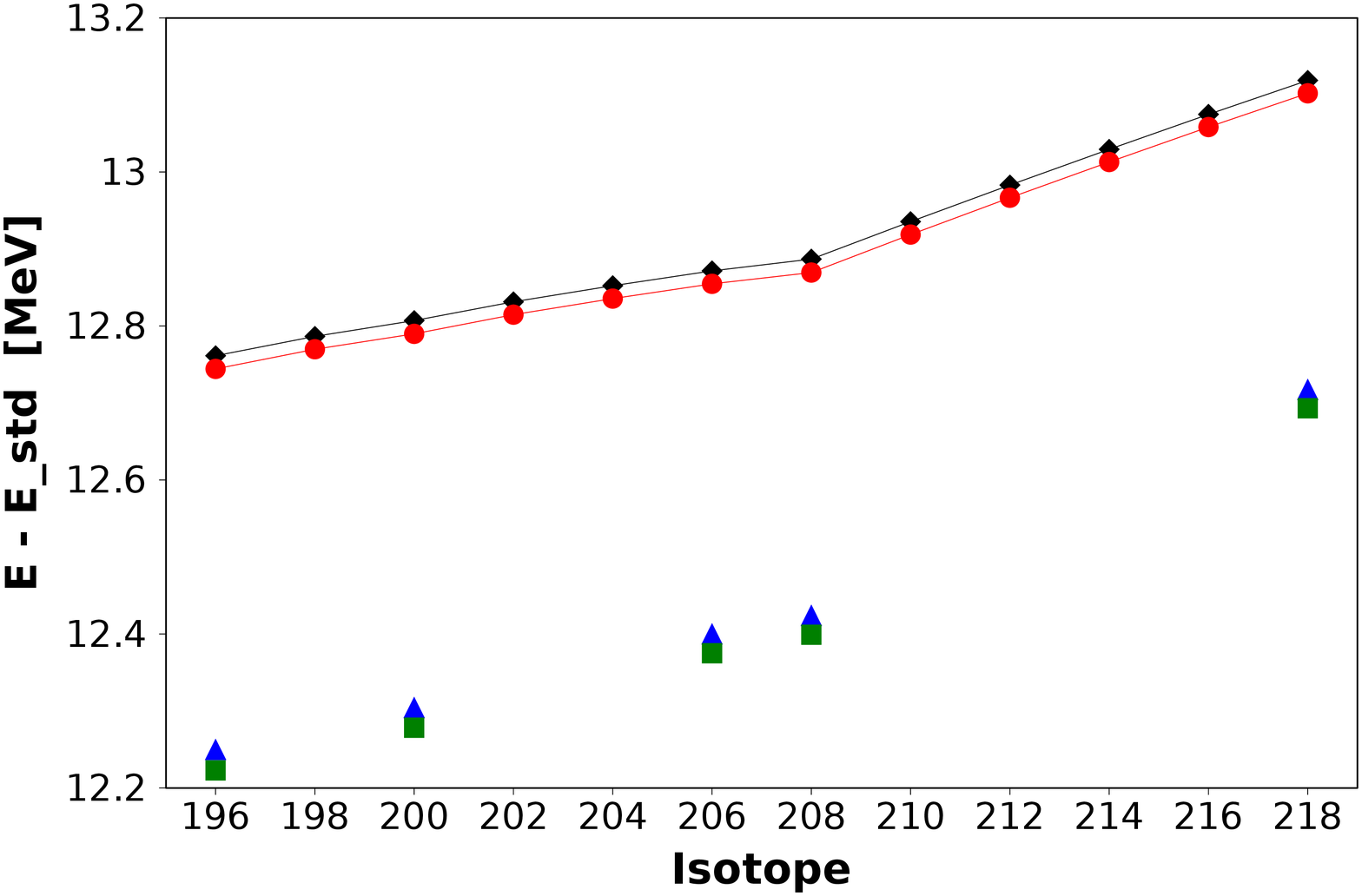}
\caption{Lipkin projected energies (\ref{Projected Energy}) relative to the 
standard SLY4 energies, calculated for the chain of lead isotopes. The results 
for the exact masses (Lipkin operator (\ref{eq:lipkin}) with $k = k_{0}$ of 
Eq. (\ref{eq:exactMass})) are given with the center-of-mass correction added 
after (diamonds) and before (circles) variation. Similarly, for the renormalized 
masses ($k$ of Eq. (\ref{eq:102})), the results for closed sub-shells are given 
with the correction added after (triangles) and before (squares) variation.}
\label{fig:lead}
\end{center}
\end{minipage}\hspace{0.03\textwidth}%
\begin{minipage}[t]{0.46\textwidth}
\begin{center}
\includegraphics[width=0.95\textwidth]{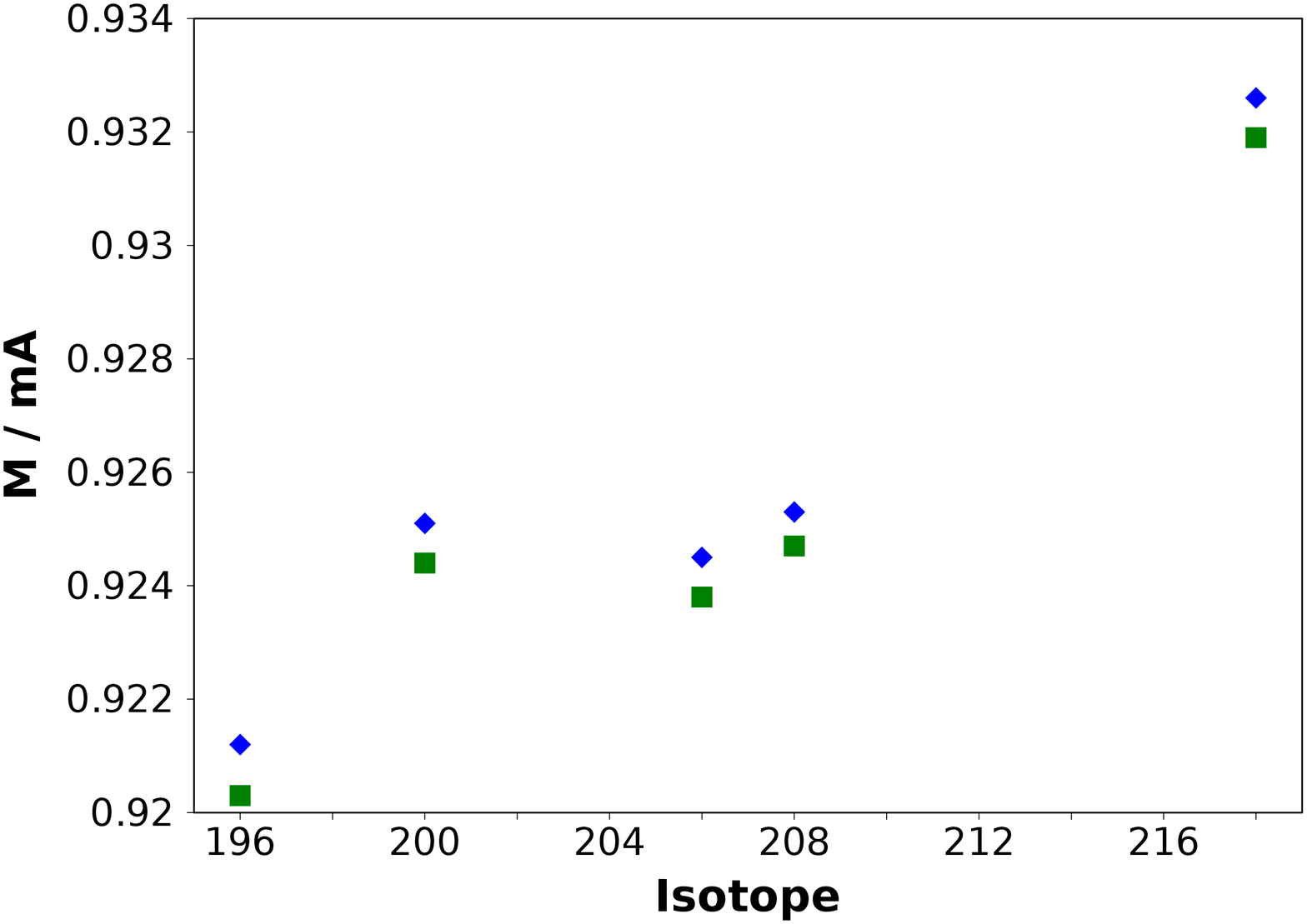}
\caption{Ratios $k_{0}/k = M/mA$ of the renormalized and exact masses determined after
(triangles) and before (squares) variation.}
\label{fig:mass}
\end{center}
\end{minipage}
\end{center}
\end{figure}

To determine the value of $k$, we proceed as follows \cite{[Lip60],[Dob09g]}. First define at each
iteration the translated wave-function $|\Phi(\mathbf{R})\rangle$ as:
\begin{equation}
|\Phi(\mathbf{R})\rangle = e^{\frac{i}{\hbar}\mathbf{R}\cdot\mathbf{\hat{P}}}
|\Phi\rangle.
\end{equation}
Then one can show that the correcting Lipkin parameter $k$ reads:
\begin{equation}
\label{eq:102}
k=\frac{h(\mathbf{R})-h(\mathbf{0})}{p_2(\mathbf{R})-p_2(\mathbf{0})},
\end{equation}
where:
\begin{equation}
\label{eq:101}
h(\mathbf{R})=\frac{\langle \Phi \mid \mathrm{\hat{H}} \mid \Phi(\mathbf{R})
\rangle}{\langle \Phi \mid \Phi(\mathbf{R}) \rangle}, \qquad
p_2(\mathbf{R})=\frac{\langle \Phi \mid \mathbf{\hat{P}}^2 \mid \Phi(\mathbf{R})
\rangle}{\langle \Phi \mid \Phi(\mathbf{R}) \rangle},
\end{equation}
are the energy and momentum kernels. The calculation of $k$ at each iteration is
therefore straightforward, since it only requires to compute $h(\mathbf{R})$ and
$p_2(\mathbf{R})$ for a single (arbitrary) value of the shift vector
$\mathbf{R}$. The parameter $k$ plays the role of a renormalized mass. It can be
compared to the traditional so-called 1-body center-of-mass correction factor:
\begin{equation}
k_{0} = \frac{\hbar^{2}}{2mA}
\label{eq:exactMass}
\end{equation}
where $m$ is the nucleon mass. Since for the translational-symmetry
restoration the Gaussian Overlap Approximation (GOA) is
excellent~\cite{[Dob09g]}, one can also approximate the Lipkin
parameter by the GOA expression~\cite{[RS80],[Dob09g]}:
\begin{equation}
\label{eq:103}
k_{\text{GOA}}=-\frac{h(\mathbf{R})-h(\mathbf{0})}
                     {4\log^2(\langle \Phi \mid \Phi(\mathbf{R}) \rangle)}\mathbf{R}^2,
\end{equation}
where one assumes that $h(\mathbf{R})$ and $\log(\langle \Phi \mid \Phi(\mathbf{R})\rangle)$
can be at the shift of $\mathbf{R}$ approximated by a parabola. The GOA
expression is much faster to evaluate, because it does not require
calculating kernels of the two-body operator $\mathbf{\hat{P}}^2$.

Figures \ref{fig:lead}--\ref{fig:mass} illustrate
various aspects of the Lipkin method for linear momentum projection. In the
captions of the figures, the term 'exact mass' refer to the quantity $k_{0}$,
and the term 'renormalized mass' to the quantity $k$.


\subsubsection{Shell Correction}

The shell-correction method relies on the Strutinsky energy theorem, which
states that the total energy $E$ of the nucleus reads:
\begin{equation}
E = E_{\text{bulk}} + \delta R_{\text{shell}},
\label{eq:strutinsky}
\end{equation}
where $E_{\text{bulk}}$ varies slowly with proton and neutron numbers, and
$\delta R_{\text{shell}}$ is a rapidly varying function of $Z$ and $N$ that
captures the quantum corrections to the liquid drop \cite{[Str67a],[Str68]}. It
was demonstrated in \cite{[Gia80]} that such a simple decomposition remains
valid when the total energy $E$ is computed microscopically as the integral of
some energy density functional or expectation value of a two-body effective
Hamiltonian at the Hartree-Fock approximation.

The shell correction must be computed from a set of s.p.\ levels
$\{e_{i} \}$, which in \pr{hfodd} are the Hartree-Fock s.p.\ energies.
In its traditional form, it is given by:
\begin{equation}
\delta R_{\text{shell}}^{(1)} = \sum_{i\in\{ occ \}} e_{i} -
\left\langle \sum_{i} e_{i} \right\rangle_{\text{smooth}},
\label{eq:oldShell}
\end{equation}
where $i\in\{ occ \}$ represents a set of occupied states (for the HF vacuum,
this is the set of the lowest $Z$ or $N$ levels), and the bracket $\langle
\dots\rangle_{\text{smooth}}$ represents the Strutinsky smoothing procedure. For
the latter, we follow the prescription presented in \cite{[Bol72fw]} and applied
on a large scale in macroscopic-microscopic calculations, e.g., in \cite{[Wer92]}.

The smoothed energy in expression (\ref{eq:oldShell}) contains a spurious contribution from
positive energy states $e_{i} > 0$ which can become problematic when approaching
the dripline. This spurious term can be removed by subtracting to Eq.
(\ref{eq:oldShell}) the smooth energy obtained for an independent gas of
particles \cite{[Kru00b],[Ver00]}. This leads to slightly different prescription
$\delta R_{\text{shell}}^{(2)}$ for the shell correction:
\begin{equation}
\delta R_{\text{shell}}^{(2)} =
\sum_{i\in\{ occ \}} e_{i}
-
\left\{
\left\langle \sum_{i} e_{i} \right\rangle_{\text{smooth}}
-
\left\langle \sum_{i} t_{i} \right\rangle_{\text{smooth}}
\right\},
\label{eq:newShell}
\end{equation}
where the $t_{i}$ are the eigenvalues of the kinetic energy operator. The shell
correction is computed twice, for protons and neutrons. For protons, the Coulomb
potential must also be taken into account by doing the substitution
$t_{i}\rightarrow (\hat{t} + \hat{V}_{\text{Cou}})_{i}$. The shell correction
(\ref{eq:newShell}) is of course evaluated at the convergence of the
self-consistent HF calculation. Both estimates $\delta R_{\text{shell}}^{(1)}$
and $\delta R_{\text{shell}}^{(2)}$ of the shell correction are available in
\pr{hfodd} and are triggered by the value of the input parameter \tv{IFSHEL}.


\setcounter{mysubsubsection}{0}

\subsection{New Numerical Features}
\label{subsec:numerical}


\subsubsection{Two-basis Method for HFB Calculations}

The two-basis method was devised in Ref.~\cite{[Gal94]} to solve the HFB
equations in spatial coordinates. The method allows for decoupling the
particle-hole and particle-particle channels from one another and using the same
technology as that developed for the HF+BCS method, even for the complete HFB
problem. The essence of the method relies on the solution of the HFB equations
in the basis of eigenstates of the particle-hole MF operator $h$. In
spatial coordinates, this operator is not diagonalized in every iteration, but a
set of eigenfunctions is evolved in imaginary time, and thus they converge to
eigenstates only at the end of the iterative process. In our case, the
self-consistent equations are solved by using the HO basis and the
imaginary-time evolution is not used; therefore, we implement the two-basis
method by explicitly diagonalizing $h$.

\begin{figure}[h]
\begin{center}
\begin{minipage}[b]{0.48\textwidth}
\begin{center}
\includegraphics[width=0.9\textwidth]{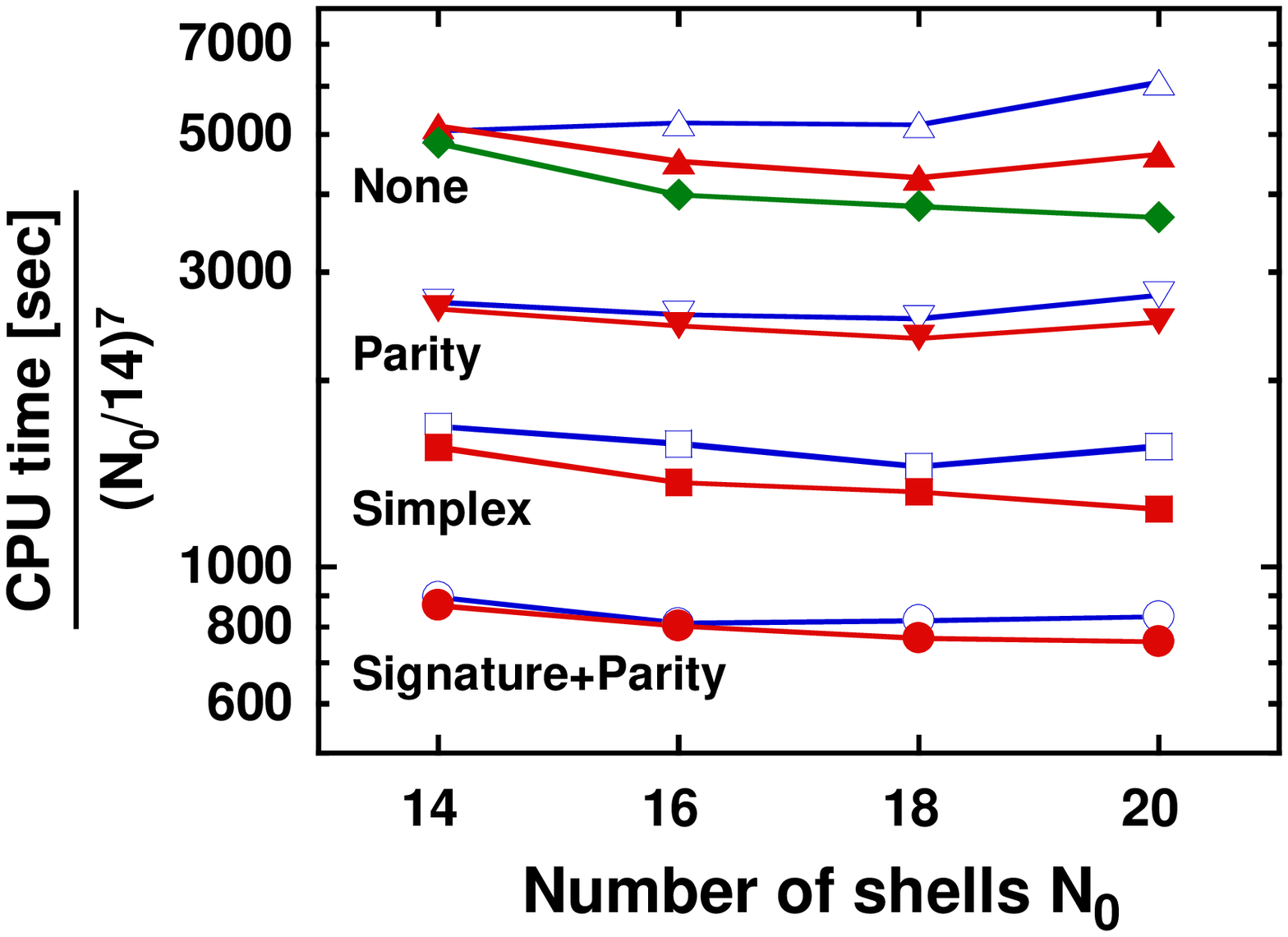}
\caption{CPU times required to perform five HFB iterations for conserved
signature and parity (circles), conserved simplex (squares), conserved parity
(down-triangles), and with no conserved symmetry (up-triangles). Standard HFB
method (open symbols) is compared with requesting the diagonalization subroutine
to return only the eigenvectors below the cutoff energy (full symbols). The
diamonds show the results obtained within the two-basis method implemented for
the case of no conserved symmetry.}
\label{fig:two.basis.CPU}
\end{center}
\end{minipage}\hspace{0.03\textwidth}%
\begin{minipage}[b]{0.48\textwidth}
\begin{center}
\includegraphics[width=0.9\textwidth]{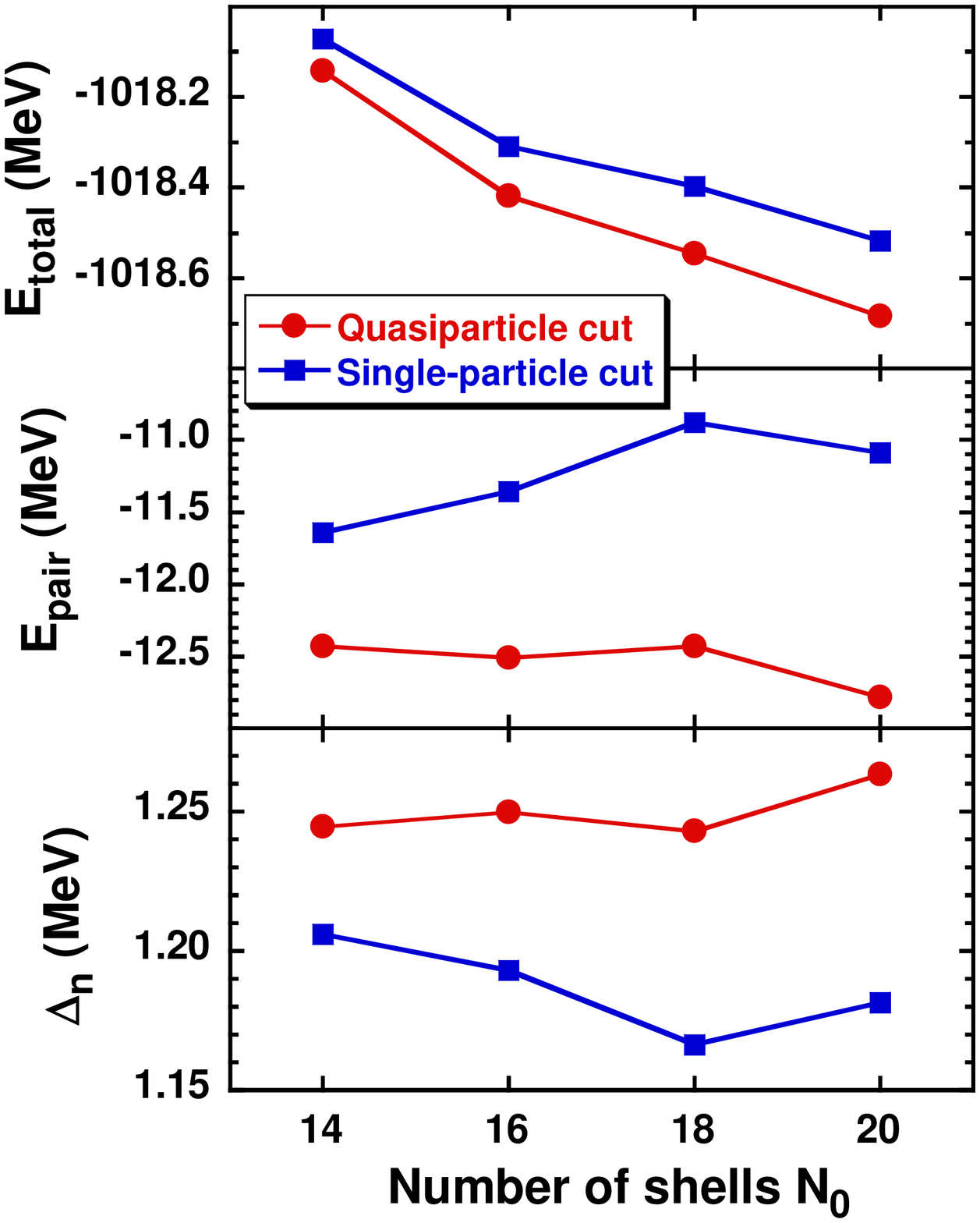}
\caption{The HFB results obtained within the standard method corresponding to the
cutoff in the quasiparticle space (circles) and within the two-basis method
corresponding to the cutoff in the s.p.\ space (squares).}
\label{fig:two.basis.results}
\end{center}
\end{minipage}
\end{center}
\end{figure}

This procedure has two advantages over the standard HFB method. First, the
cutoff of the configuration space can now be performed in the s.p.\
space and not in the quasiparticle space. Therefore, the dimension of the HFB
equations, reduced to s.p.\ states with energies $\epsilon$ below the
cutoff energy,  $\epsilon \le \bar{e}_{\text{max}}$, is much smaller than the
full HO space. This speeds up the calculations. Second, HFB calculations in a
reduced s.p.\ space do not suffer from formal drawbacks related to
the cutoff in the quasiparticle space, see discussion in Ref.~\cite{[Dob05f]}.

Figure~\ref{fig:two.basis.CPU} shows the CPU times required to perform the HFB 
calculations for $^{120}$Sn by using the cutoff energy of 
$E_{\text{cut}}=60$\,MeV, Skyrme functional SLY4, and pairing-force 
parameters of Eq.~(14) in Ref. [4], $V_0=-285.88$\,MeV, $\rho_0=0.32$\,fm$^{-3}$, 
and $\alpha=1$. The results were obtained for bases of $N_0=14$--20 and for 
four conserved-symmetry conditions, as indicated in the Figure. The CPU times 
scale as $N_0^7$. It turns out that in the case of calculations performed 
without any conserved symmetry, the two-basis method can be up to 30\% faster 
than the standard HFB method. However, almost half of this gain can be obtained 
by simply requesting in the standard HFB method that the HFB wave functions be 
calculated only below the quasiparticle cutoff energy (see keyword 
\tk{CUT\_SPECTR}). In view of this limited gain in the CPU time, in version 
(v\codeversion) the two-basis method is not implemented in the remaining three 
conserved-symmetry conditions.

The two-basis method gives results that are close, but not identical, to those
given by the standard HFB method. This is illustrated in
Figure~\ref{fig:two.basis.results}, where the total energies (upper panel) can
differ up to 200\,keV, the pairing energies (middle panel) up to 1.5\,MeV, and
the neutron pairing gaps (bottom panel) up to 70\,keV. Although these
differences are non-negligible, they are probably inferior to other
uncertainties of the approach, and at present the physical advantages or
disadvantages of one method over the other cannot be established. One should
stress that the two methods of implementing the cutoff give exactly the same
numbers of quasiparticles, so the above difference are not caused by different
sizes of the model spaces.


\subsubsection{Augmented Lagrangian Method}

Multi-constrained EDF calculations are a crucial ingredient of a number of
nuclear structure applications.  The microscopic description of high-spin states
is based on the cranking model, which requires a constraint on the value of the
total angular momentum.  Modeling the fission process involves the calculation of
multi-dimensional potential energy surfaces, where the constraints are imposed
on expectation values of the multipole moments.  Most generally,
beyond-mean-field applications, or multi-reference EDF, rely on a basis of
constrained MF states used to generate collective motion.

In previous versions of \pr{hfodd}, constraints on the nuclear shape took
the standard quadratic form, see Eq. (22) in \cite{[Dob97c]}. This so-called
quadratic penalty method was chosen to avoid the pitfalls of the method of
Lagrange multipliers (linear constraints), which often fails to converge. The
quadratic penalty method is usually very fast and robust, but does not always
yield the desired solution: the expectation value of the multipole moments at
convergence may differ, sometimes significantly, from the requested values. In
addition, it depends rather sensitively on the stiffness parameter, which controls
the magnitude of the corrective term.

The Augmented Lagrangian Method (ALM) provides a valuable alternative for
multi-const\-rained calculations \cite{[Hes69],[Pow69]}. It is a general
algorithm which aims at minimizing a scalar function $\mathcal{E}(\gras{x})$ of
the vector $\gras{x}$ under a set of constraints $g_{i}(\gras{x}) = q_{i}^{0}$
(the so-called finite-dimensional, equality-constrained nonlinear optimization problem).
In practice, it can simply be viewed as a smart combination of both the linear and
quadratic penalty methods. Adopting the same notations as in Sec.\ 2.3 of
\cite{[Dob97c]}, the total energy takes the form:
\begin{equation}
\mathcal{E}' = \mathcal{E}
- \sum_{\lambda\mu}L_{\lambda\mu}
( \langle \hat{Q}_{\lambda\mu} \rangle - \bar{Q}_{\lambda\mu})
+ \sum_{\lambda\mu}C_{\lambda\mu}
( \langle \hat{Q}_{\lambda\mu} \rangle - \bar{Q}_{\lambda\mu})^{2},
\label{eq:totalE}
\end{equation}
where $L_{\lambda\mu}$ is the Lagrange parameter for the multipole
$(\lambda,\mu)$, $C_{\lambda\mu}$ is the corresponding stiffness and
$\bar{Q}_{\lambda\mu}$ the requested value for the multipole moment
$\hat{Q}_{\lambda\mu}$. Note that the minus sign for the linear constraint term
is a matter of convention. While the stiffness is an input parameter that remains
constant along the iterations, $L_{\lambda\mu}$ has to be re-adjusted. At
 iteration $m+1$, the new Lagrange parameter reads:
\begin{equation}
L_{\lambda\mu}^{(m+1)} = L_{\lambda\mu}^{(m)}
- 2C_{\lambda\mu} ( \langle \hat{Q}_{\lambda\mu} \rangle^{(m)}
- \bar{Q}_{\lambda\mu}).
\label{eq:update}
\end{equation}

For a EDF solver already implementing the quadratic penalty method, adding the
ALM is extremely simple: (i) The linear term in Eq. (\ref{eq:totalE}) must be
added to the total energy, (ii) the matrix elements of the corresponding HF
potential $U^{(m)} = - \sum_{\lambda\mu} L_{\lambda\mu}^{(m)}
\hat{Q}_{\lambda\mu}$ must be added to the HF(B) matrix, and (iii) the Lagrange
parameter must be updated at every iteration according to Eq. (\ref{eq:update}).
The method induces almost no computational overhead, is very robust, and always
gives very precisely the requested solution, see \cite{[Sta10]}.


\subsubsection{Linear Constraints Based on the RPA Matrix}
\label{sec:rpa}

The ALM method does not make any specific hypotheses as to how the function
$\mathcal{E}(\gras{x})$ is computed. Within the nuclear EDF, the function
$\mathcal{E}(\gras{x})$ is the total energy of the nucleus, and is itself obtained as
the solution to a variational problem. One may therefore take advantage of this
additional information to adapt the standard optimization algorithm with linear
constraints. Such an approach was proposed already 30 years ago in the context
of the self-consistent nuclear MF theory with finite range effective forces
\cite{[Dec80]}. At every iteration, an estimate of the QRPA matrix at the cranking
approximation is computed. This information is used to make an educated update
of the Lagrange parameters $L_{\lambda\mu}$ of the linear constraints.
A detailed and pedagogical presentation of the algorithm and its applications for
fission barriers calculations can be found in \cite{[You09]}.

The implementation of the method in \pr{hfodd} follows very closely the Appendix
A of \cite{[You09]}, and we refer to this work for further information. Let us note 
that this method requires the matrix of the constraint operator in the q.p. basis, 
which involves 8 matrix multiplications per iteration (4 for neutrons, 4 for protons). 
The computation of the constraints correlation matrix also requires 
$N_{\text{c}}^{2}$ additional matrix multiplications per iteration, where $N_{\text{c}}$ 
is the number of constraints. When simplex symmetry is conserved, the properties of 
the basis in \pr{hfodd} reduce the size of the matrices involved in all these operations 
to one half of the total basis size at most. The computation overhead can still 
be noticeable, but is always largely compensated by a drastic reduction in the number 
of iterations necessary to reach convergence and the near-perfect precision 
of the obtained solution. All major linear algebra operations (matrix multiplication 
and inversion) are carried out by BLAS and LAPACK routines.


\subsubsection{Interface with \pr{hfbtho}}

The code \pr{hfbtho} solves the Skyrme HFB equations in the HO basis by assuming
axial and reflection symmetry \cite{[Sto05]}. These built-in symmetries make the
HFB matrix block-diagonal, and the typical run time of the program is at least an
order of magnitude shorter than for \pr{hfodd}. This makes \pr{hfbtho} an ideal
tool for large-scale calculations in cases where axial- and reflection
symmetries are sensible assumptions \cite{[Sto09a]}. Conversely, \pr{hfbtho} is
not appropriate for specific problems such as the description of nuclear
fission, where the complexity of the nuclear shapes impose the use of a fully
symmetry-unrestricted solver like \pr{hfodd}.

Both codes have been carefully benchmarked against one another in even-even
\cite{[Dob04a]} and odd nuclei \cite{[Sch10]} at the equal-filling
approximation. The difference of total energies in a nucleus like $^{120}$Sn is
typically of the order of 10 eV, and can entirely be attributed to the different
techniques of computing the Coulomb potential. Such a nearly perfect match makes
it possible to accelerate \pr{hfodd} run time by coupling the two codes together:
for a given nucleus, the calculation is first carried out by \pr{hfbtho}
(assuming axial and reflection symmetry), then restarted with \pr{hfodd} after a
simple unitary transformation. If the physical solution is axial and parity
invariant, the \pr{hfbtho} solution is the correct one, and \pr{hfodd} can stop
after the basis transformation. If the solution breaks any of these symmetries,
the self-consistent procedure will continue until convergence. The motivation
for such an interface is the observation that, for nearly all nuclear shapes,
the driving deformation is the axial quadrupole moment $Q_{20}$.

Let us denote by
$\{|\text{SIM}_{n}\rangle\} \equiv \{ |n_{x},n_{y},n_{z}; s=\pm i\rangle \}$ the
simplex-conserving Cartesian Harmonic Oscillator basis used in \pr{hfodd}. We
denote by $\{|\text{CYL}_{n}\rangle\}$ the cylindrical harmonic oscillator basis
used in \pr{hfbtho},
$|\text{CYL}_{n}\rangle \equiv |N, n_{\rho}, n_{z}, \Lambda, \Omega\rangle$. The
basis transformation
$\{|\text{CYL}_{n}\rangle\} \rightarrow  \{|\text{SIM}_{n}\rangle\}$
proceeds in two steps:
\begin{enumerate}
\item The coordinate transformation $\{|\text{CYL}_{n}\rangle\}\! \rightarrow\!
\{|\text{CAR}_{n}\rangle\}$ is carried out, where the
$\{|\text{CAR}_{n}\rangle\}$ $\equiv |n_{x},n_{y},n_{z}; \sigma\rangle$ are the
Cartesian harmonic oscillator states and $\sigma = \pm 1/2$ is the $z$-projection
of the intrinsic spin;
\item A unitary phase transformation is then performed to go to the good
y-simplex basis: $\{|\text{CAR}_{n}\rangle\} \rightarrow \{|\text{SIM}_{n}\rangle\}$.
\end{enumerate}

{\bf Coordinate transformation  -} The spatial quantum numbers in Cartesian and
cylindrical coordinates are related through:
\begin{equation}
N = n_{x} + n_{y} + n_{z} = 2n_{\rho} + \Lambda + n_{z}.
\end{equation}
Let us note $n_{\perp} = N - n_{z}$. In principle:
\begin{equation}
0 \leq n_{\rho} \leq n_{\perp}, \ \ \text{and} \ \ \
-n_{\perp} \leq \Lambda \leq +n_{\perp}.
\end{equation}
However, quantum numbers $\Lambda < 0$ (therefore $n_{\rho} > n_{\perp}/2$)
correspond to states which are the time-reversed partners of the states $\Lambda
> 0$: In \pr{hfbtho}, time-reversal symmetry is conserved, all basis states with
$\Lambda < 0$ are disregarded, and the HFB matrix $\mathcal{H}$ is explicitly
block-diagonal by $\Omega = \Lambda \pm \sigma$ values. The full HFB matrix is
therefore reconstructed from each $\Omega$-block. By a suitable reordering of
indexes, it is then split into 4 matrices $\mathcal{H}^{(\sigma\sigma')}$. The
$\mathcal{H}^{(\sigma\sigma')}$ are purely spatial matrices with matrix elements
of the type:
\begin{equation}
\langle n'_{\rho} \Lambda' n'_{z} | \hat{\mathcal{H}}^{(\sigma\sigma')} |
n_{\rho} \Lambda n_{z}\rangle.
\label{eq:mat_el}
\end{equation}

\begin{figure}[h]
\centering
\includegraphics[width=0.6\textwidth,angle=-90]{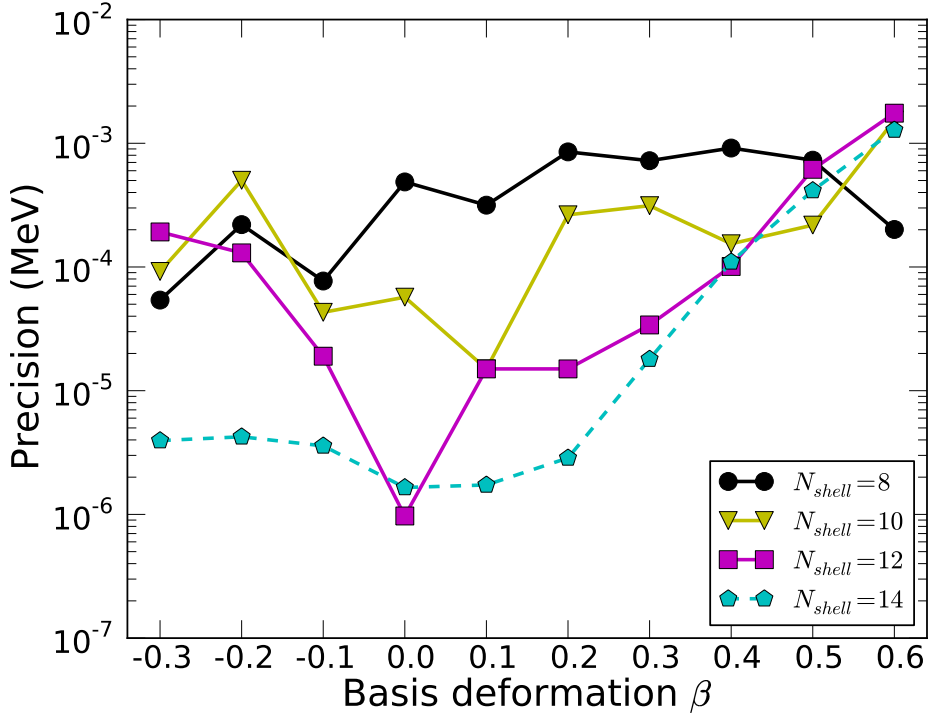}
\caption{Stability of the \pr{hfodd} solution after restart from \pr{hfbtho} at
the spherical point in $^{152}$Dy, as function of the deformation $\beta_{2}$ of
the HO basis.}
\label{fig:interface}
\end{figure}

The transformation of the matrix elements (\ref{eq:mat_el}) from cylindrical to
Cartesian coordinates requires the overlaps: $\langle n_{x} n_{y} n_{z} |
n_{\rho} \Lambda n_{z}\rangle$. We use the formulas given in \cite{[Dav91]}. Let
us recall:
\begin{multline}
\langle n_{x} n_{y} n_{z} | n_{\rho} \Lambda n_{z}\rangle
=
\delta_{2n_{\rho}+\Lambda,n_{x}+n_{y}}(-i)^{n_{y}} (-1)^{\Lambda}
\left[ \frac{n_{x}!n_{y}!n_{z}!(n_{\rho}+\Lambda)!}{2^{2n_{\rho}+\Lambda}}
\right]^{1/2} \\
\times\sum_{p=p_{\text{min}}}^{p_{\text{max}}}
\frac{(-1)^{p}}{p!(n_{x}-p)!(p+n_{\rho}-n_{x})!(n_{\rho}+\Lambda-p)!},
\end{multline}
with:
\begin{equation}
p_{\text{min}} = \left\{
\begin{array}{ll}
0              & \text{for}\ n_{\rho} \geq n_{x}, \\
n_{x}-n_{\rho} & \text{for}\ n_{\rho} \leq n_{x}, \\
\end{array}
\right.
\end{equation}
and:
\begin{equation}
p_{\text{max}} = \left\{
\begin{array}{ll}
n_{x}            & \text{for}\ n_{x} \leq n_{\rho}+\Lambda, \\
n_{\rho}+\Lambda & \text{for}\ n_{x} \geq n_{\rho}+\Lambda. \\
\end{array}
\right.
\end{equation}
The overlaps for $\Lambda <0$ are obtained from those with $\Lambda >0$
according to:
\begin{equation}
\langle n_{x} n_{y} n_{z} | n_{\rho} \Lambda_{(<0)} n_{z}\rangle
=
(-1)^{|\Lambda|}
\langle n_{x} n_{y} n_{z} | n_{\rho} \Lambda_{(>0)} n_{z}\rangle^{*}.
\end{equation}

{\bf Phase transformation  -} After the matrices $\mathcal{H}^{(\sigma\sigma')}$
are obtained in Cartesian coordinates, the phase transformation, Eqs.
(78a)-(78b) of Ref. \cite{[Dob97c]}, is performed. Let us recall that it reads:
\begin{equation}
\begin{array}{l}
|n_{x} n_{y} n_{z}, s=+i\rangle =
\displaystyle\frac{1}{\sqrt{2}} \left(
i^{n_{y}}|n_{x} n_{y} n_{z}, \sigma=\frac{1}{2}\rangle - i^{-n_{y}+1}|n_{x} n_{y}
n_{z},\sigma=-\frac{1}{2}\rangle
 \right), \medskip \\
|n_{x} n_{y} n_{z}, s=-i\rangle =
\displaystyle\frac{1}{\sqrt{2}} \left(
-i^{n_{y}+1}|n_{x} n_{y} n_{z}, \sigma=\frac{1}{2}\rangle + i^{-n_{y}}|n_{x} n_{y}
n_{z},\sigma=-\frac{1}{2}\rangle
 \right),
\end{array}
\label{eq:phase}
\end{equation}
where $s = \pm i$ is the y-simplex. In \pr{hfbtho}, simplex and time-reversal
symmetries are always conserved (by construction), so that the HFB matrix is
block diagonal: $\mathcal{H}^{(ss')} = \delta_{ss'}\mathcal{H}^{(s)}$ and
$\mathcal{H}^{(-s)} =\mathcal{H}^{(s)*}$. The blocks $\mathcal{H}^{(s=\pm i)}$
are obtained by linear combinations of the $\mathcal{H}^{(\sigma\sigma')}$ and
the phase factors recalled in Eq.(\ref{eq:phase}).

For axially- and parity-symmetric HFB solutions, the interface between
\pr{hfbtho} and \pr{hfodd} gives a precision at restart that ranges from 1 eV to
1 keV depending on the nucleus, the characteristics of the basis and the
quadrupole deformation of the solution. Almost all the error is contained in the
direct Coulomb energy which is computed differently in the two codes. Figure
\ref{fig:interface} shows the stability of the \pr{hfodd} iterations immediately
after restart from the \pr{hfbtho} solution at the spherical point in
$^{152}$Dy, for different deformations of the basis and different basis sizes.


\subsubsection{Mixing of Matrix Elements of the HFB Matrix}

When solved by successive diagonalizations, as in \pr{hfodd}, the Hartree-Fock
equations are self-consistent. In practice, the iterative scheme is started with
a set of initial conditions, formally some vector $\gras{V}^{(0)}$, that
linearize the problem. Upon entering iteration $m$ with a vector
$\gras{V}_{\text{in}}^{(m)}$, solving the HF equations yields a new vector
$\gras{V}_{\text{out}}^{(m)}$. This vector is then used as input to the next
iteration $m+1$, $\gras{V}_{\text{in}}^{(m+1)} \rightarrow
\gras{V}_{\text{out}}^{(m)}$. The iterations stop when
$|\gras{V}_{\text{out}}^{(m+1)} - \gras{V}_{\text{out}}^{(m)}| \leq
\varepsilon$, with $\varepsilon$ a measure of the convergence. In practice, the
input vector at iteration $m+1$ must be a mixing of $\gras{V}_{\text{in}}^{(m)}$
and $\gras{V}_{\text{out}}^{(m)}$ for the iterations to converge. This mixing
can be a simple linear mixing of the type:
\begin{equation}
\gras{V}_{\text{in}}^{(m+1)} = \alpha\gras{V}_{\text{out}}^{(m)} + (1 - \alpha)
\gras{V}_{\text{in}}^{(m)}
\end{equation}
or more elaborate such as produced by the Broyden method \cite{[Bar08]}.

Both the linear and Broyden mixing are implemented in \pr{hfodd}. By default,
the iterated quantities $\gras{V}$ are the values of the HF fields on the
Gauss-Hermite integration mesh \cite{[Dob09d]}. However, it was noticed that
when the Lipkin-Nogami (LN) prescription is used, the matrix elements of the
density matrix in the HO basis must also be added in order to ensure
convergence. In the simplest case where time-reversal and simplex symmetries are
conserved, and the LN procedure is applied to both protons and neutrons, the
size of the Broyden vector is augmented by $4M^{2}$, where $M$ is the size of
the s.p. basis. For large bases, the size of the Broyden vector can thus become
prohibitive. To by-pass this memory bottleneck, the mixing of the matrix
elements of the HFB matrix has been implemented.

In \pr{hfodd}, the self-consistent loop is organized in such a way that at
each iteration $m$, it is initialized with the set of HF fields (for neutrons
and protons) $\gras{V}_{\text{in}}^{(m)}$, and ends with the determination of
the new fields $\gras{V}_{\text{out}}^{(m)}$: it therefore lends itself
naturally to mixing the HF fields. By contrast, the mixing of the matrix
elements of the HFB matrix is most easily performed when the self-consistent
loop starts with an initial HFB matrix $\gras{H}_{\text{in}}^{(m)}$ and ends
with the computation of the new HFB matrix $\gras{H}_{\text{out}}^{(m)}$
(this is the case, e.g., in \pr{hfbtho}). In order to conserve the 'HF
potential-oriented' structure of the self-consistent loop of \pr{hfodd}, the
mixing of the matrix elements of the HFB matrix must be done immediately
after a new $\gras{H}_{\text{out}}^{(m)}$ has been calculated: in \pr{hfodd}
such a condition actually requires independent mixing for protons
and neutrons with two separate calls to the mixing routine and, in the case of the
Broyden mixing, two different memory arrays.

\begin{figure}[h]
\centering
\includegraphics[width=0.6\textwidth,angle=-90]{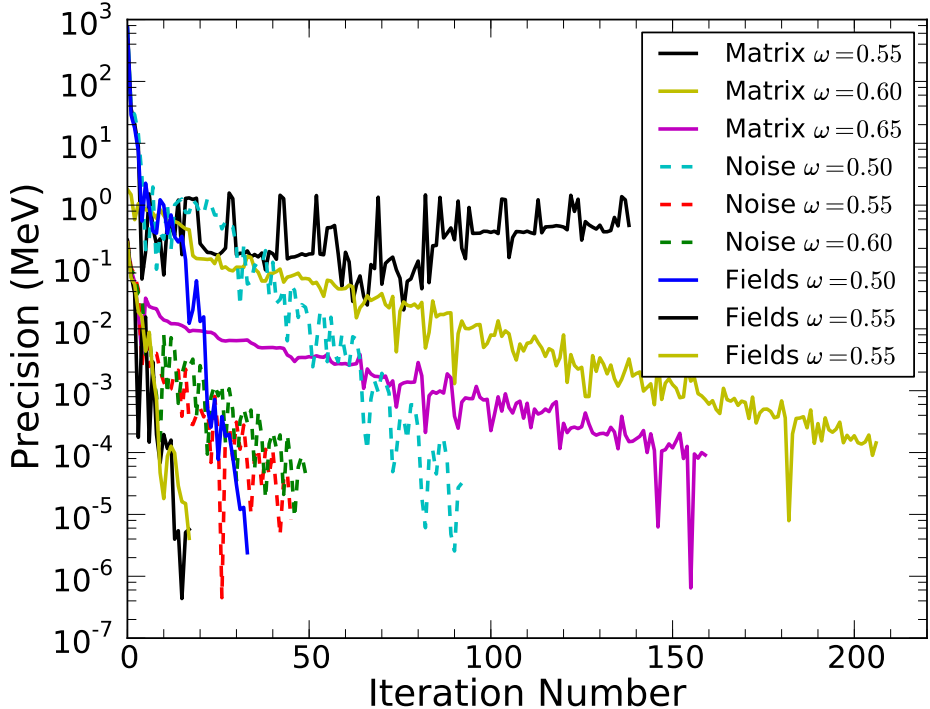}
\caption{Convergence of the Cranking HF iterations in $^{151}$Tb with the
Broyden method. Curves labeled 'Matrix' correspond to the original mixing of
matrix elements; curves marked 'Noise' to the mixing of matrix elements when the
Broyden memory is erased every 4 iterations; curves marked 'Fields' correspond
to the mixing of HF fields on the Gauss-Hermite mesh.}
\label{fig:mixing}
\end{figure}

The size of the Broyden vector (for one isospin only) depends on the symmetries
of the problem: simplex conserved {\tt ISIMPY=1} or broken {\tt ISIMPY=0},
time-reversal symmetry conserved {\tt IROTAT=0} or broken {\tt IROTAT=1}, HFB
calculations {\tt IPAHFB=1} or HF calculations {\tt IPAHFB=0}. Taking also into
account the symmetries of the matrix of the mean field and pairing field, the
size of the Broyden vector is:
\begin{multline}
N = \text{\tt (1+IROTAT)}\times  M^{2}
  + \text{\tt (1-ISIMPY)}\times  M^{2} \\
  + \text{\tt IPAHFB} \left[ M(M+1)/2 + \text{\tt IROTAT}\times M(M-1)/2 \right]
\end{multline}
For a typical static HFB calculation with conserved time-reversal and simplex
symmetry, a stretched basis such that {\tt NXHERM}={\tt NYHERM}=30, {\tt
NXHERM}=60 and $M = 1000$, and 7 iterations conserved in memory, the Broyden
method requires 302 MB of RAM when fields are mixed, and about 168 MB when
matrix elements are mixed.

It has been observed that the mixing of the matrix elements of the HFB matrix in
\pr{hfodd} is less stable than in a comparable implementation in \pr{hfbtho}.
This numerical noise may be due to the fact that \pr{hfodd} breaks a number of
symmetries that are conserved in \pr{hfbtho}, and which manifest themselves by
numerically small, non-zero elements in the matrix. Correct performance for ground-state
calculations can still be obtained if the memory of the Broyden method is erased
every $n$ iterations, with $n < n_{\text{mem}}$ where $n_{\text{mem}}$ is the
number of iterations retained to compute the full Broyden correction (noise
cancellation). In practice $n_{\text{mem}} = 4$ gives decent results.


\setcounter{mysubsubsection}{0}

\subsection{Parallel Programming Model}
\label{subsec:parallel}

Starting in version (v\codeversion), the code \pr{hfodd} has built-in parallel
capabilities. These capabilities are controlled by 3 different pre-processor
options and are discussed below.


\subsubsection{Distributed Memory Parallelism}
\label{subsubsec:mpi}

Density functional theory is an efficient method to compute the properties of
multi-fermionic systems. From a programming point of view, recasting all degrees
of freedom of the problem into a single function of $\gras{r}$, the local
one-body density matrix, allows the formulation of a compact, CPU- and memory-thrifty,
implementation. In practice, the average computation time of standard nuclear
EDF solvers ranges from a few seconds for a spherical closed-shell nucleus up to
a few hours for a full symmetry-breaking configuration in a heavy nucleus.

Such naive estimates, however, are deceptive: in many instances, the nature of
the problem at hand requires the computation of {\it many} such configurations,
as for example in the determination of Potential Energy Surfaces (PES), which
are critical ingredients in the proper description of the nuclear fission
process. While the time of calculation of any point of the N-dimensional PES may
be of the order of a few hours, systematic and accurate mapping of the surface
is required to compute reliable estimates of barrier heights, tunneling
probabilities or collective inertia. For $N=5$ degrees of freedom with $n_{i} =
10$ sample points each, the size of the mesh is 100,000: such a problem requires
both supercomputers and an adapted programming model.

One giant simplification of high-performance computing applications with EDF
methods is that the theory generates by construction a naturally parallel
computational problem: most of the time, all configurations can be handled
independently by a single core (CPU unit) of a multi-core processor. The amount
of inter-processor communication is therefore often rather low (coarse granularity).
Such a property has made it possible to compute the entire mass table in less
than a day \cite{[Sto09a]}.

The distributed memory programming model of \pr{hfodd} contains two layers of
parallelism managed by the standard Message Passing Interface (MPI) library. The
outermost layer corresponds to $N_{\text{master}}$ master groups of cores,
each group being in charge of computing a given nuclear configuration. The
innermost or group layer is made of the $N_{\text{proc}}$ cores in any given
group. The division of the processor grid in these two layers is carried out at
the very beginning of the code using standard MPI group and communicator
routines. Most applications of \pr{hfodd} do not require the
group structure, that is, $N_{\text{proc}} = 1$ is sufficient most of the time.
Examples of distributed HFB calculations are discussed in Sec.
\ref{subsubsec:scalapack}.

From a user's point of view, running \pr{hfodd} on several cores requires the
following:
\begin{itemize}
\item The code must be compiled by setting the pre-processor option
\tv{USE\_MPI}=1. The user is in charge of ensuring that an implementation of MPI
exists on his/her system.
\item Input data now falls into 2 categories: process-dependent and process-independent
data, the word 'process' referring to a given HFB calculation.
Process-independent data is everything that will be the same on every process.
Contrariwise, process-dependent data is what changes from one process to the next: it
is therefore what distinguishes the nuclear configurations ($Z$, $N$, constraints,
etc.). For practical reasons, see Section \ref{subsubsec:input}, process-independent
data will {\it always} be contained in the file \tv{hfodd.d}, and the process-dependent
data {\it always} in the file \tv{hfodd\_mpiio.d}.
\end{itemize}


\subsubsection{Scalable Input Routines}
\label{subsubsec:input}

In its single core version, \pr{hfodd} reads its input data from the system
standard input. In practice, data is often contained in a simple ASCII text
file, and the execution of the code uses input redirection. Input is read by the
routine \tv{NAMELI}, which loops over the file for specific keywords, each
keyword being immediately followed by the relevant data. Such a structure
provides good flexibility, since the user can add or remove keywords at will
from the input file.

In parallel mode, duplicating this structure of the I/O operations on all
available cores is not efficient, and can in fact affect the stability of
the entire system. Indeed, not only would all cores try to access the disks
more or less at the same time, but they would all access the {\it same} file and
seek different positions in it. It is a well-known rule of thumb that parallel I/O
has to be handled either by one core only, or using dedicated libraries. In
the specific case of \pr{hfodd}, the amount of input data is rather small,
common to and needed by all cores, and read only at the beginning of the
calculation. The most natural solution is therefore to assign one core to
the task of reading it and broadcasting it to the others.

However, the flexibility induced by the keyword structure of the input file now
becomes a disadvantage, since the amount of data to be read (and then
broadcast) is in fact known only at run time. This feature suggests the use of
Fortran 90 linked lists for each basic type of input data (integer, double
precision real number and character string). The entire I/O operation is then
broken down in 3 phases: (i) construction of the linked list (ii) broadcast and
(iii) reconstruction of local data. In the first phase, the routine
\tv{mpi\_getSequentialData()} parses the file \tv{hfodd.d}, which has exactly
the same structure as the standard \pr{hfodd} input file, and for each data,
adds a node to the relevant linked list. At the end of the process, linked lists
are copied to local allocatable arrays which are then broadcast to all
cores using the MPI routine \tv{MPI\_Bcast} (second phase). All cores
then acquire a local (to a given core) copy of 3 arrays, for the 3 basic
types of data mentioned above. Finally, every element of these
arrays is re-associated with the relevant local \pr{hfodd} variables (third phase). This is done
by the routine \tv{mpi\_setSequentialData}.

\begin{figure}[h]
\begin{center}
\begin{minipage}[t]{0.48\textwidth}
\centering
\includegraphics[width=0.7\textwidth,angle=-90]{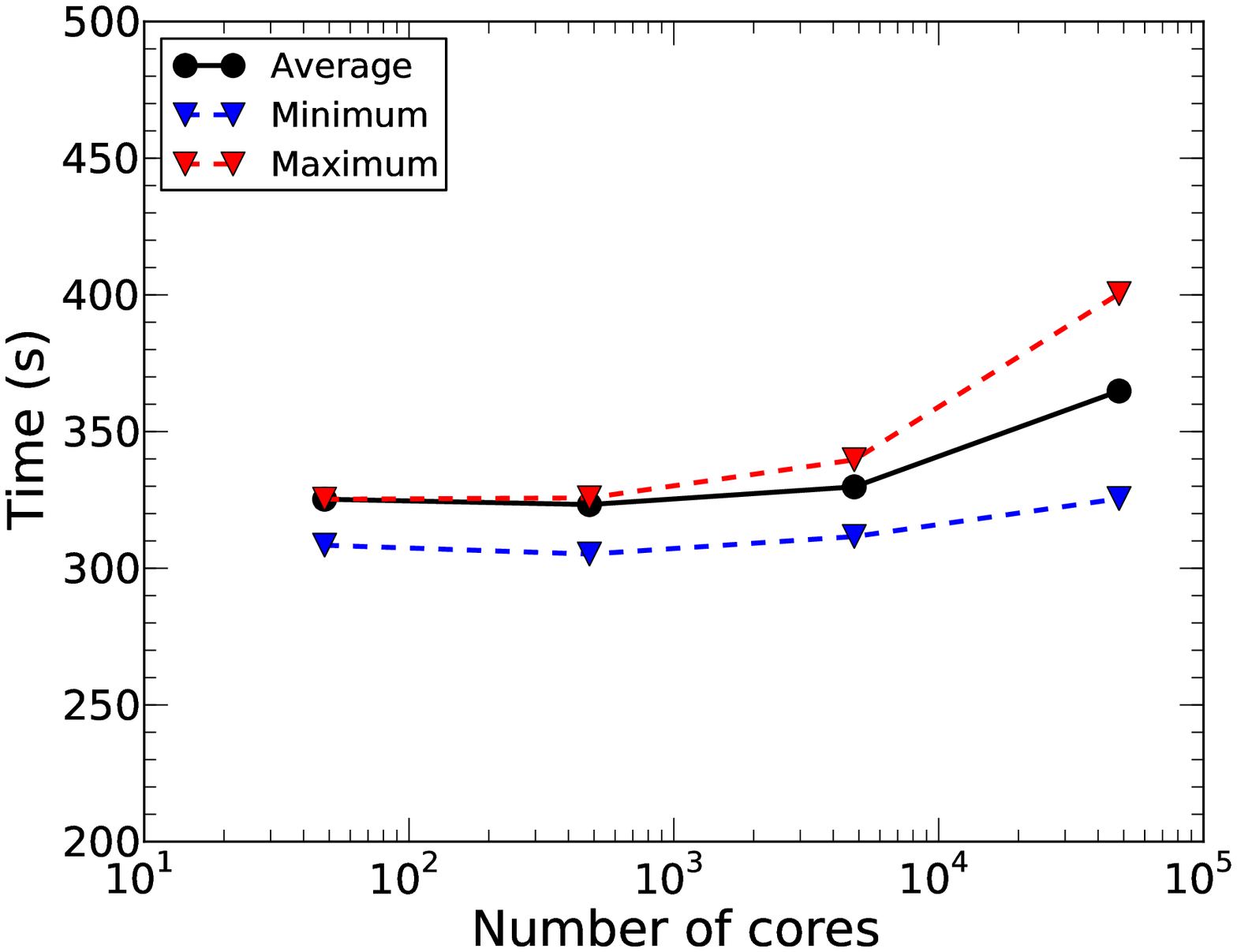}
\caption{Average, minimum and maximum CPU time for 6 HFB iterations in
$^{152}$Dy as a function of the number of cores for a full spherical HO basis of
$N_{\text{shell}} = 14$ shells. The pairing cut-off energy is $E_{\text{cut}}=60$
MeV. Calculations were done on the Cray XT5 at NCCS, the code was compiled
with the PGI v.10.3 compiler with \tv{-fast -Mipa=fast} options. The Cray
\tv{libsci} library was used for BLAS and LAPACK. All cores do the same HFB
calculation.}
\label{fig:scaling}
\end{minipage}\hspace{0.03\textwidth}%
\begin{minipage}[t]{0.48\textwidth}
\centering
\includegraphics[width=0.7\textwidth,angle=-90]{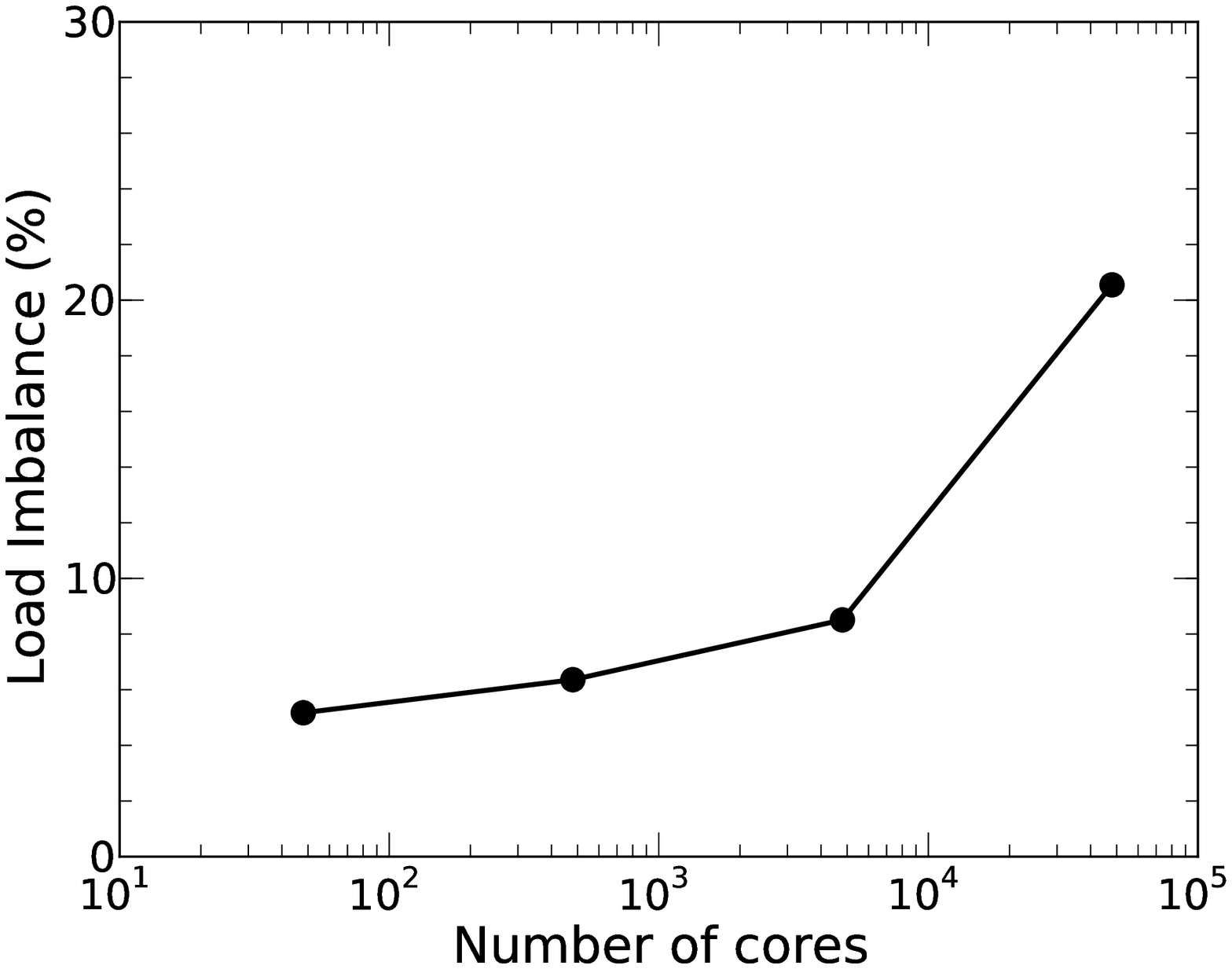}
\caption{Load imbalance for the same runs as in Fig. \ref{fig:scaling}. Load
imbalance is defined here as $(T_{\text{max}} - T_{\text{min}}) / T_{\text{avr}}$,
where $T_{\text{min}}$ (resp. $T_{\text{max}}$) is the minimum (resp. maximum)
time of execution over all cores, and $T_{\text{avr}}$ the average time over all cores.}
\label{fig:imbalance}
\end{minipage}
\end{center}
\end{figure}

In principle, it would have been enough to implement this scheme for the
standard \pr{hfodd} input file, and add a few keywords relevant for
process-dependent data. However, it proved more convenient to put all
process-dependent data in a file of its own, with a similar keyword structure.
The I/O process described above has therefore to be repeated for the
process-dependent data. This is carried out by the routines
\tv{mpi\_getParallelData()} and \tv{mpi\_setParallelData}.

This implementation of the I/O procedure combines a number of advantages. First
and paramount, it scales well with the number of cores available, since
only one core is dedicated to accessing the disk and reading the data.
Since the amount of data will always be very small (a few kB at most), the
broadcast phase should not induce a very large communication overhead. In
addition, the linked list structure conserves the flexibility to add/remove data
from the input files, which also ensures backward and forward compatibility with
all future releases of the code.

Figure \ref{fig:scaling} shows the scaling properties of \pr{hfodd}. The sample
calculation consisted of 6 HFB iterations in $^{152}$Dy in a full spherical basis
of $N = 14$ shells with constraints on $Q_{20} = 20$ b and $Q_{22} = 0$ b. All
cores computed the same configuration. The PGI compiler v10.3 with the \tv{-fast
-Mipa=fast} options was used to compile the code, the Cray library \tv{libsci} to
link to BLAS and LAPACK, and all calculations were done on the Jaguar Cray XT5
at the NCCS. These technical characteristics are given because the actual time of
execution can vary very significantly depending on the compiler/platform and
compiler options used.

Since all cores perform exactly the same calculation, any departure from a flat
straight line should be attributed at first order to (i) the inter-core communication during the
initial input setup (ii) filesystem operations to create/access files. While the input
setup has been somewhat optimized, see above, all other I/O operations are
the same as in serial mode: for the runs shown in
Figs. (\ref{fig:scaling}-\ref{fig:imbalance}), every core generates 2 files that
remain opened with read/write permissions for the entire time of execution. To
better grasp the impact of the lack of I/O optimization, figure \ref{fig:imbalance}
shows the load-imbalance of this calculation, defined here as:
\begin{equation}
\text{LI} = \frac{T_{\text{max}} - T_{\text{min}} }{T_{\text{avr}}}
\end{equation}
where $T_{\text{min}}$ (resp. $T_{\text{max}}$) is the minimum (resp. maximum)
time of execution over all cores, and $T_{\text{avr}}$ the average time over all
cores. Perfect load-balancing (equal distribution of work between cores)
implies LI=0.


\subsubsection{Shared Memory Parallelism}
\label{subsubsec:openmp}

Many leadership class computers have adopted a hybrid distributed-shared memory
architecture, whereby all cores are grouped into processors, each processor having
access to its own physical memory. Tests of the current version of \pr{hfodd} have been
carried out on the Franklin Cray XT4 and Jaguar Cray XT5, which are
characterized by, respectively, 4-core processors with 8 GB shared memory and 12-core
processors with 24 GB shared memory. As mentioned earlier, most applications of
\pr{hfodd} can run on a single core, so that in the case of the Jaguar
supercomputer, 12 simultaneous  calculations can be run on a processor. However,
for large basis size, the memory needed by one instance of \pr{hfodd} can exceed
the 2 GB/core available.

\begin{figure}[h]
\begin{center}
\begin{minipage}[t]{0.48\textwidth}
\centering
\includegraphics[width=0.95\textwidth]{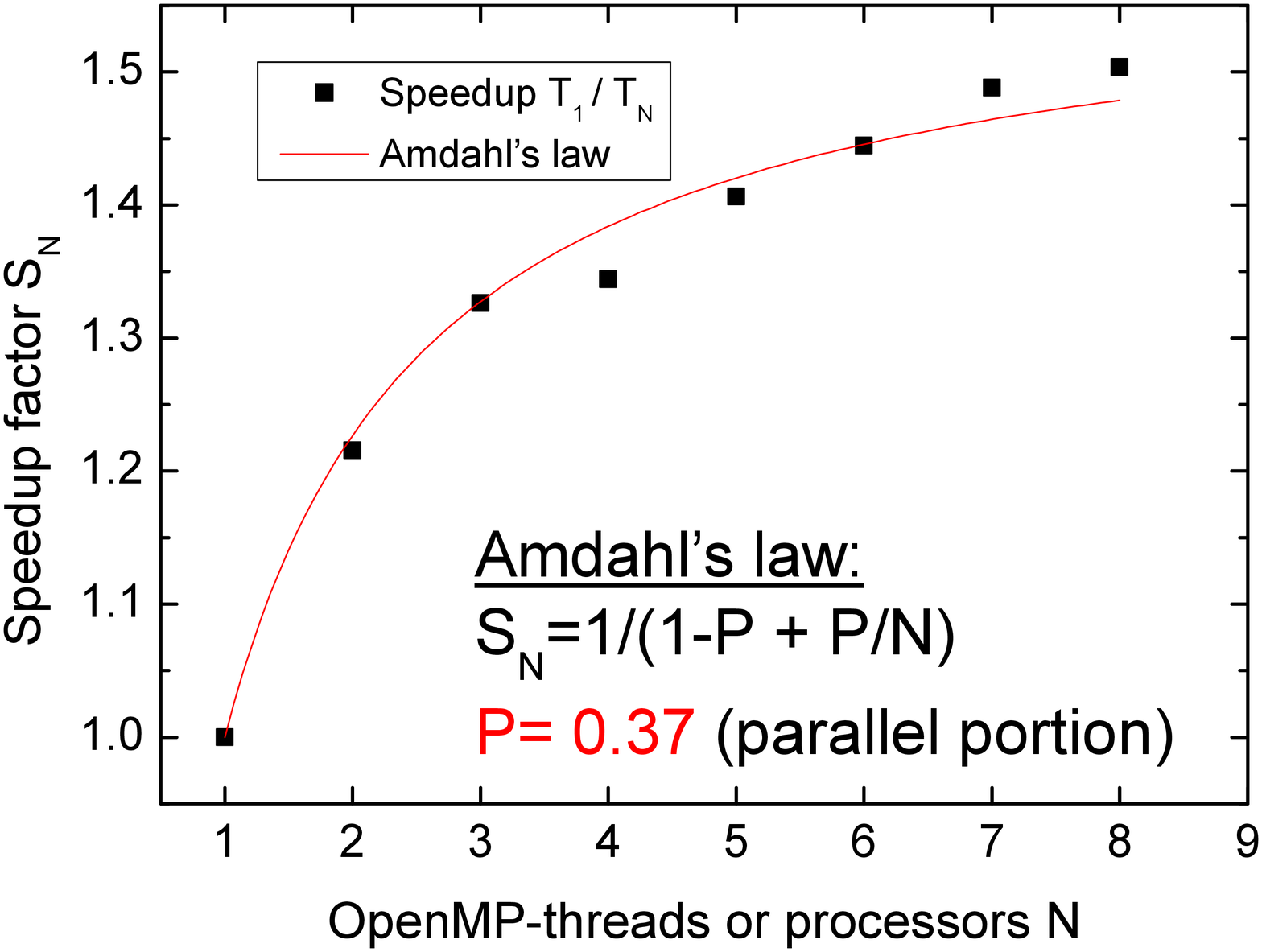}
\caption{Multi-threading accelerations in HFB calculations of $^{180}$Hg
for a deformed $N_{\text{shell}} = 26$ HO basis with 1140 states and deformation
$\alpha_{20} = -0.32$. Calculations were performed on a Intel Xeon processor with the
Intel Fortran Compiler and the \tv{-xSSE4.2 -O3 -override-limits} options with
standard BLAS libraries, and are compared to Amdahl's law. }
\label{fig:omp_threads1}
\end{minipage}\hspace{0.03\textwidth}%
\begin{minipage}[t]{0.485\textwidth}
\centering
\includegraphics[width=1.0\textwidth]{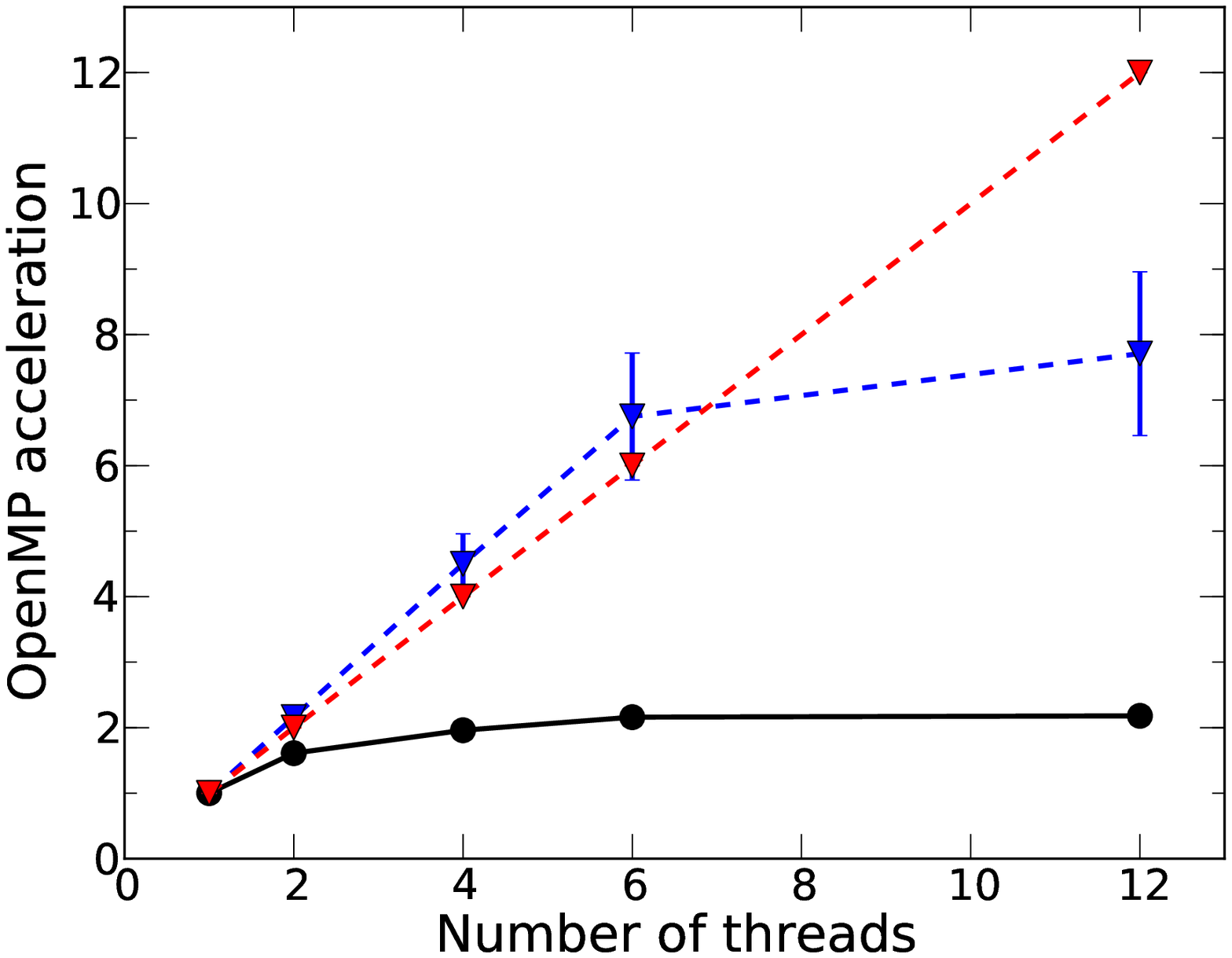}
\caption{Same as Fig. \ref{fig:omp_threads1} for 6 HFB iterations in $^{152}$Dy in
a full spherical basis of $N = 14$ shells (same characteristics as in
Fig. \ref{fig:scaling}) and threaded BLAS libraries. Black circles: speed-up of the
entire code; triangles: speed-up for the 3 impacted routines (error bars reflect the
rounding of the time to the nearest second). Dashed line: perfect scaling for the
OpenMP acceleration.}
\label{fig:omp_threads2}
\end{minipage}
\end{center}
\end{figure}

This seemingly limitation of the prevailing architectures can be exploited to
accelerate the execution of the code by using the standard OpenMP API. If the
number of instances of the code running on a given processor is less than the actual
number of cores in that processor, several cores are in fact inactive. The OpenMP API
offers a very simple interface to recycle these cores for quasi-automatic
parallelization of the code (multi-threading). The execution then consists of
series of sequential instructions (on one core) combined with multi-core
parallel sequences. This mechanism is particularly effective to
(quasi-)automatically parallelize loops. Let us recall that OpenMP instructions,
or pragmas, are coded in the form of comments only activated by the relevant
option of the compiler: modifications of the source code remain minimal.

We identified 3 subroutines of \pr{hfodd} that could take a significant chunk of
the total run time: subroutine \pr{DENMAC} computes the density matrix from the
HFB eigenvectors; subroutine \pr{SPAVER}  computes s.p.\ expectation
values of operators; subroutine \pr{NILABS} defines the Nilsson labels of
s.p.\ states. All these routines involve 3-nested loops with
$N_{\text{states}}$ elements, where $N_{\text{states}}$ is the number of basis
states. For large bases with $N_{\text{states}} \approx 1000-2000$, these loops
become time-consuming, and they cannot be re-arranged easily
in a way that memory access is optimized.

Figure \ref{fig:omp_threads1} shows the OpenMP acceleration of a full HFB calculation
in $^{180}$Hg with a large basis of 26 HO shells and 1140 states (basis
deformation $\alpha_{20} = -0.32$) as function of the number of threads. Calculations
were performed on a cluster of Intel Xeon processors with the Intel Fortran
Compiler and the \tv{-xSSE4.2 -O3 -override-limits} options. Standard
(un-threaded) BLAS and LAPACK libraries were used. The 3 routines impacted
by Open MP pragmas represent a small fraction $P = 0.37$ of the total execution
time in this case, and the observed acceleration nicely aligns with predictions by
the empirical Amdahl's model.

Leadership class computers often provide threaded BLAS libraries. In Fig.
\ref{fig:omp_threads2}, we shows the acceleration induced by OpenMP in the
same test case as in Sec. \ref{subsubsec:input}.  This profiling experiment was
done on the Jaguar Cray XT5 computer at the NCCS by linking to threaded BLAS
libraries. At the level of the 3 modified routines, the scaling is perfect with
the number of threads; overall acceleration is a little better than in the case of
Fig. \ref{fig:omp_threads1} due to the benefit of using threaded libraries. OpenMP
acceleration is activated by setting \tv{USE\_OPENMP} to 1 in the Makefile.


\subsubsection{Parallelization of Diagonalization Routines}
\label{subsubsec:scalapack}

\begin{figure}[h]
\begin{center}
\begin{minipage}[t]{0.48\textwidth}
\centering
\includegraphics[width=0.7\textwidth,angle=-90]{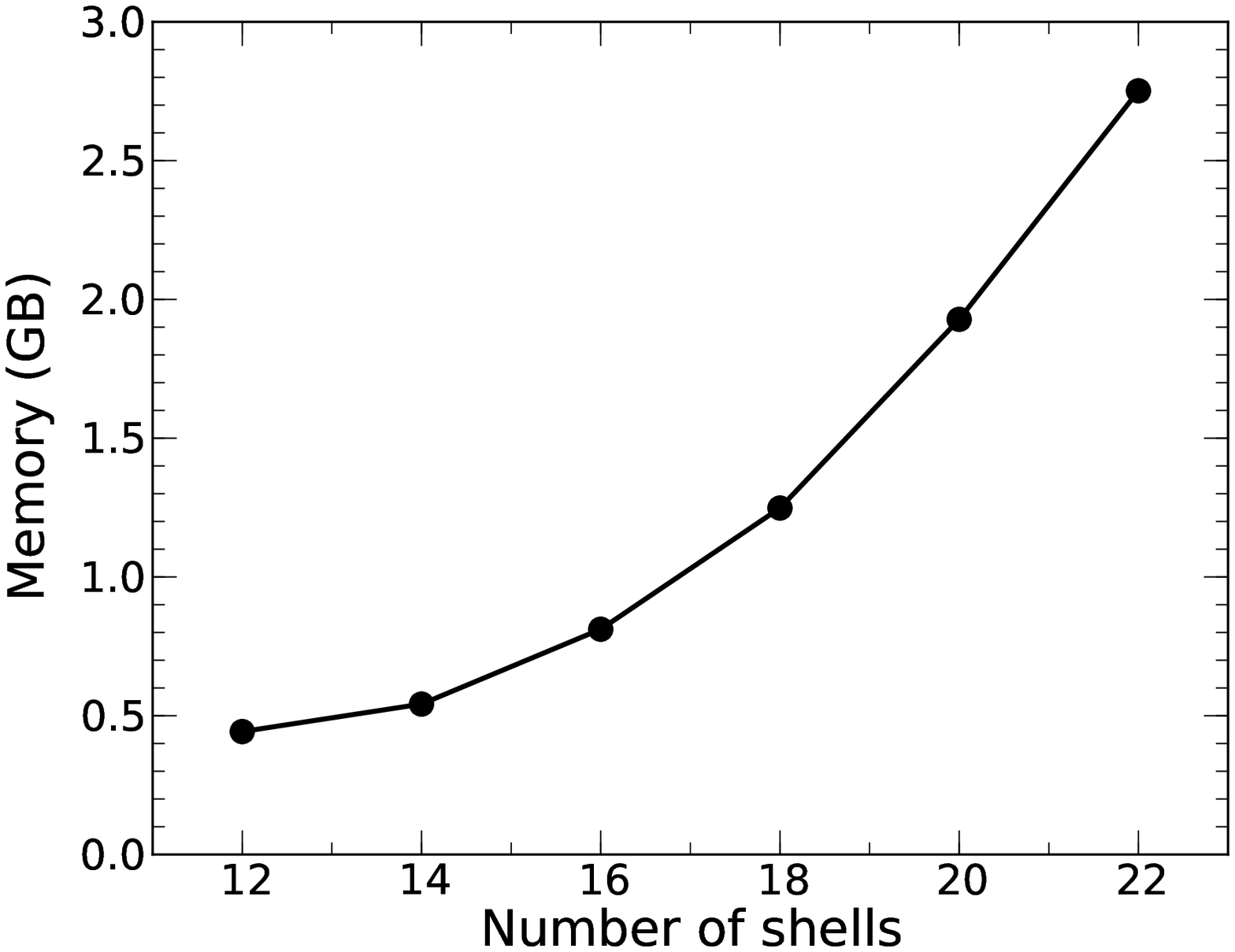}
\caption{The memory high-water mark attained by \pr{hfodd} is plotted against
the number of major harmonic oscillator shells used in the basis.  }
\label{fig:profile_peakmem}
\end{minipage}\hspace{0.03\textwidth}%
\begin{minipage}[t]{0.48\textwidth}
\centering
\includegraphics[width=0.7\textwidth,angle=-90]{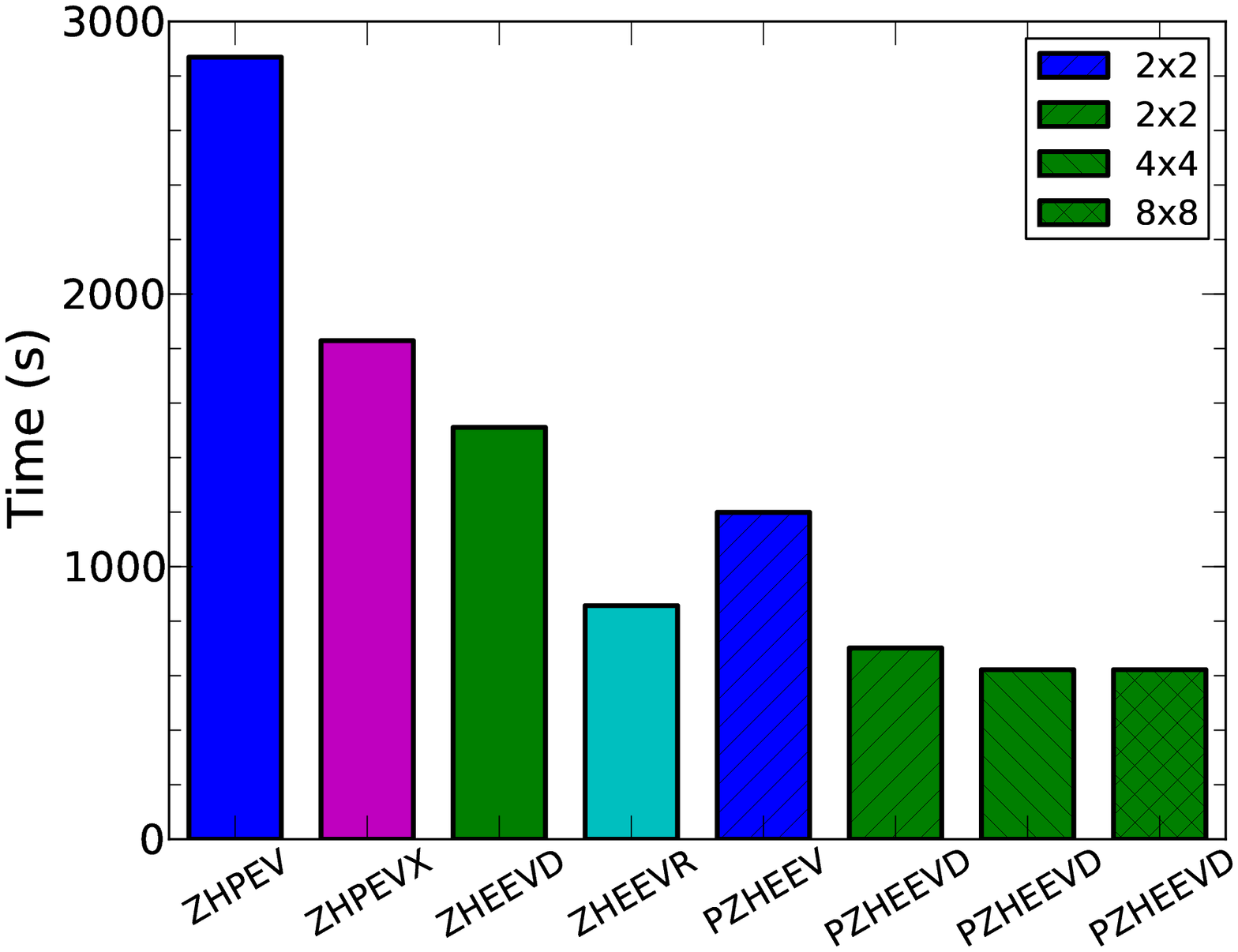}
\caption{Total execution time of \pr{hfodd} spent for different diagonalization
routines and numbers of rows used in a square ScaLAPACK process grid. For
single-core tests, \tv{HFBISZ} was modified so that all eigenstates of the HFB
matrix are computed, to allow meaningful comparisons with multi-core algorithms.}
\label{fig:profile_timing}
\end{minipage}
\end{center}
\end{figure}

One strength of the \pr{hfodd} code is its ability to perform computations
without assuming symmetries.  Calculations without symmetries, however, prove
computationally expensive.  Until now \pr{hfodd} has been successfully used in
several massively parallel applications, such as the survey of one
quasi-particle states in odd mass nuclei \cite{[Sch10]} and potential energy
surfaces for fission \cite{[Sta09]}.  The trend in massively parallel computing,
however, is to reduce the memory available to each computing core, thereby
indirectly imposing restrictions on the symmetries assumed in current-generation
\pr{hfodd} calculations.  In Fig.~\ref{fig:profile_peakmem}, the peak memory
used in a \pr{hfodd} run is plotted against the number of full shells in the
harmonic oscillator basis.  It is worth noting that the present NCCS Cray XT4
and Cray XT5 machines have $2$GB available to a core, which is exceeded by a
problem using $20$ full oscillator shells.  For problems like fission that require
the calculation of highly deformed nuclei, at least $26-31$ oscillator shells
can be needed.

To fully take advantage of the next generation of massively parallel platforms
as well as its full symmetry-breaking capabilities, \pr{hfodd} should therefore
be scaled so that its solution of the HFB equations is distributed among an
arbitrary number of cores. Before undertaking such a daunting task, though,
the anticipated benefits must be assessed and demonstrated. In this release of
the code, we have tested whether we can achieve any gain in the speed of the
diagonalization routines by using the ScaLAPACK library.

Tests were conducted for the all symmetry-breaking HFB case (subroutine
\tv{HFBSIZ}), which involves the largest matrices, of size $4N \times 4N$ if
$N$ is the number of states in the HO basis. Fig.~\ref{fig:profile_timing} shows
the time of execution of 3 HFB iterations in $^{152}$Dy in a full spherical HO
basis of $N_{\text{shell}} = 14$ shells ($N=680$ states). The code was compiled
with the \tv{-fast -Mipa=fast} option and run on the Cray XT5 computer at
the NCCS. OpenMP acceleration was activated, and the BLAS and LAPACK
routines were provided by the \tv{libsci} library. Fig. \ref{fig:profile_timing}
shows the benchmark results for different diagonalization methods and
different ScaLAPACK core grids. Let us note that the ScaLAPACK library
does not include a parallelized version of all LAPACK diagonalization routines.
From the strict perspective of execution time, the single-core version of
\pr{hfodd} invoking the \tv{ZHEEVR} LAPACK routine (based on a very fast
diagonalization algorithm) may turn out to be faster than the multi-core
version, which can only resort to \tv{PZHEEV} or \tv{PZHEEVD} ScaLAPACK
methods.

The module \tv{hfodd\_sl} provides an interface to the parallel matrix
diagonalization routines available in ScaLAPACK.  The user may enable the
use of the \tv{hfodd\_sl} module by setting the environment variable
\tv{USE\_SCALAPACK} to $1$ in the project Makefile (similarly,
\tv{USE\_SCALAPACK} is set to $0$ to disable the module).  The
\tv{hfodd\_sl} module requires that the program is linked to a ScaLAPACK
library and compiled with MPI, as well. The user should also specify the size
of the process grid in the project Makefile through the environment variables
\tv{M\_GRID} (number of process rows) and \tv{N\_GRID} (number of process
columns).  The program will stop with an error, if the product \tv{M\_GRID
$\times$ N\_GRID} times the number of different HFB configurations 
exceeds the number of processes allocated for the program.

\setcounter{mysubsubsection}{0}

\subsection{Corrected errors}
\label{subsec:bugs}

In the present version (v\codeversion), we have corrected the
following little significant errors of the previous published version
(v2.40h)~\cite{[Dob09d]}.

\subsubsection{Coulomb energies}
\label{subsubsec:coulomb}

For the combination of the Coulomb parameters {\tv{ICOUDI}}=1 and
{\tv{ICOUEX}}=2 ({\tv{ICOUDI}}=2 and {\tv{ICOUEX}}=1), the direct
(exchange) Coulomb energy was inadvertently set to zero.

\subsubsection{Skyrme parameters}
\label{subsubsec:Skyrme}

For SLY5 and SLY7, the predefined values of the Skyrme-force parameters
corresponded to those given in Ref.~\cite{[Cha95a]}, while for SLY4 they
corresponded to unrounded numbers communicated by the authors of the force.
At present, all these parameters are coded according to Ref.~\cite{[Cha98]},
while the previous values are kept in the code under acronyms
sLy4, sLy5, and sLy7.

\subsubsection{Quasiparticle blocking}
\label{subsubsec:blocking}

The time reversal of s.p.\ states was incorrectly coded for
{\tv{IDSIGB}} = $-$1 (in {\ts{BLOSIG}}),    {\tv{IDSIMB}} = $-$1 (in {\ts{BLOSIM}})
{\tv{IDSIQB}} = $-$1 (in {\ts{BLOSIQ}}),    {\tv{IDSIZB}} = $-$1 (in {\ts{BLOSIZ}}).
Moreover, the quasiparticle blocking was incorrectly performed in {\ts{BLOSIZ}}
and, in some situations, these errors compensated one another. One should
stress that the calculated results were always correctly converged,
but they could have corresponded to other blocked quasiparticles as
compared to what was requested in the input data.

\subsubsection{Yukawa mean fields}
\label{subsubsec:Yukawa}

For calculations related to the Yukawa interaction that were not
supposed to be using the mean fields stored on the disc
({\tv{IFCONT}} = 0), in the first iteration the Yukawa mean fields
were inadvertently set to zero. Since later these mean fields were
calculated correctly, the error was only affecting the continuation of
calculations from the disc, while the converged results were correct.

\subsubsection{The Broyden method}
\label{subsubsec:Broyden}

In the subroutine {\ts{DOBROY}}, the input parameter {\tv{ALPHAM}} was 
inadvertently reset to 1$-${\tv{SLOWEV}} and thus it was ineffective.


\section{Input Data File}
\label{sec:input}


\subsection{Physics}

\key{PROJECTISO} 0, 2, 1, 1.E-6, 0, 0 =
				\tv{IPRGCM}, \tv{ISOSAD}, \tv{NBTKNO}, \tv{EPSISO}, \tv{ICSKIP},
\tv{IFERME}

\noindent For \tv{IPRGCM}$\geq$ 1 and \tv{NBTKNO}$\geq$ 1, the isospin projection is carried out.
The isospin projection is performed for all values of isospin $T$
such that $T_{\text{min}} \leq T \leq T_{\text{min}} + \Delta T$. In the current
implementation, $T_{\text{min}} = (N-Z)/2$, and $\Delta T$ := \tv{ISOSAD}/2.
The number of Gauss-Legendre points required to
perform integrations over the Euler angle $\beta_{T}$ is given by \tv{NBTKNO}.
By setting \tv{ICSKIP}=1, in the projection routines the Coulomb
interaction can be switched off, that is, in Eq.~(\ref{ham})
$\hat{V}^{C}$ can be neglected. Parameter
\tv{EPSISO} gives $\varepsilon_{T}$ and controls the number of
good-isospin states. Parameter
\tv{IFERME} controls the calculation of the
Fermi matrix element (\ref{fermi}), which proceeds in two independent
runs. In the first run, for \tv{IFERME}=$-$1, the wave-function $|I=0,
T\approx 1, T_{z}=\pm 1\rangle $ is computed and stored in the
external file under the name specified in the keyword
\tv{WAVEF-FILE}. Next, for \tv{IFERME}=+1, the matrix element
(\ref{fermi}) is computed. This run uses information on the $|I=0,
T\approx 1, T_{z}=\pm 1\rangle $ state calculated in the first step
and supplied in the file specified again in the keyword
\tv{WAVEF-FILE}.

Current restrictions: In version (v\codeversion), the isospin projection is only available at the
Hartree-Fock level, that is, it requires \tv{IPAIRI}=0. When the full Hamiltonian is
re-diagonalized (\tv{ICSKIP}=0), the Coulomb potential must be computed
exactly, that is, by expanding both the direct and exchange terms onto a sum of
Gaussians, see Sec.~2.10 of Ref.~\cite{[Dob09]}. This requires setting \tv{ICOUDI}=2
and \tv{ICOUEX}=2 in the input. The method also imposes setting \tv{IROTAT}
to 1.  Note also that \tv{IPRGCM}$\geq$ 1 activates either the isospin-, or
angular-momentum projection, or both (see keyword \tk{PROJECTGCM} described in
\cite{[Dob09]}). To run the isospin- or angular-momentum
projection alone, one needs to set the numbers of integration points \tv{NUAKNO}=1
and \tv{NUBKNO}=1 or \tv{NBTKNO}=1, respectively.

\key{COULOCHARG} 1.0 =
                                \tv{E\_EFFE}

\noindent \tv{E\_EFFE} is the factor that multiplies the value of the
elementary charge used in the Coulomb mean-field. In this way the
strength of the Coulomb interaction can be modified or, for
\tv{E\_EFFE}=0, the Coulomb interaction can be switched off. Note
that this factor does not change the strength of the Coulomb interaction
rediagonalized within the isospin-projection method, see
Sec.~\ref{subsubsec:isospin}.

\key{FINITETEMP} 0.0 =
				\tv{TEMP\_T}

\noindent This keyword controls the value of the nuclear temperature (in MeV).
If \tv{TEMP\_T}$>$0, The finite-temperature HFB or HF+BCS calculations are
performed.

\key{SHELLCORCT} 0 =
				\tv{IFSHEL}

\noindent For \tv{IFSHEL}=1, the traditional shell correction $\delta
R_{\text{shell}}^{(1)}$ is computed at convergence from the HF s.p.\
energies. If \tv{IFSHEL}=2, the shell correction $\delta R_{\text{shell}}^{(2)}$
is computed, which includes the removal of spurious contributions from positive
energy states. For \tv{IFSHEL}=0, shell correction is not computed.

\key{SHELLPARAM} 1.2, 1.2, 4.5, 6 =
				\tv{GSTRUN}, \tv{GSTRUP}, \tv{HOMFAC}, \tv{NPOLYN}

\noindent This item adjusts the parameters of the shell correction. The
variable \tv{GSTRUN} (\tv{GSTRUP}) stands for the $\gamma_{\text{n}}$
($\gamma_{\text{p}})$ smoothing parameter for the neutrons (protons).
\tv{HOMFAC} is the multiplicative factor $\alpha$ that defines the energy window
for the shell correction, according to $\alpha \hbar\omega$, with $\hbar\omega =
41/A^{1/3}$. \tv{NPOLYN} is the number $p$ of Hermite polynomials used in the
expansion of the smooth density. Preset values are a good choice to use for 
\tv{IFSHEL}=1. For \tv{IFSHEL}=2, the recommended
values are $\gamma_{\text{n}} = 1.54, \gamma_{\text{p}} = 1.66$, $\alpha = 4.5$
and $p = 10$.

\key{RENORMASS} 0, 0.0, 0.0, 0.0 =
                    \tv{IRENMA}, \tv{DISTAX}, \tv{DISTAY}, \tv{DISTAZ}

\noindent For \tv{IRENMA} $\ge$ 1 the renormalized translational mass is
determined in each iteration by using components of the shift vector $\mathbf{R}$,
\tv{DISTAX}, \tv{DISTAY}, and \tv{DISTAZ} (in fm), in the $x$, $y$,
and $z$ direction, respectively, multiplied by \tv{IRENMA}, see
Eqs.~(\ref{eq:101}) and (\ref{eq:103}). For \tv{IRENMA}=0, the mass
is not renormalized.

\key{GAUOVERAPP} 1 = \tv{IDOGOA}

\noindent For \tv{IDOGOA}=0 or 1, the translational mass is
determined by using the Lipkin method (\ref{eq:101}) or the GOA
expression (\ref{eq:103}), respectively. However, even for \tv{IDOGOA}=0
the GOA mass is calculated and printed for reference. For
\tv{IRENMA}=0, the value of \tv{IDOGOA} is ignored.

\key{UNEDF\_PROJ} 0 = \tv{IF\_EDF}

\noindent Setting \tv{IF\_EDF}=1 activates specific parameterizations of the 
Skyrme energy functional for which volume coupling constants are determined 
automatically from nuclear matter properties (used as inputs) and surface 
coupling constants are preset as usual, following the strategy laid out in 
\cite{[Kor10a]}.


\subsection{Numerical Methods}

\key{TWOBASIS} 0 =
				\tv{ITWOBA}

\noindent For \tv{ITWOBA}=1, the two-basis method is used to
diagonalize the HFB Hamiltonian. \tv{ITWOBA}=1 requires
\tv{IPAHFB}=1. The two-basis method is currently implemented only
for the no-symmetry case; therefore, \tv{ITWOBA}=1 requires
\tv{ISIMPY}=0 and \tv{IPARTY}=0.

\key{CUT\_SPECTR} 0 =
                                \tv{LIMQUA}

\noindent For \tv{LIMQUA}=1, the HFB quasiparticle energies are
calculated only up to the cut-off energy of \tv{ECUTOF}, see Sec.~3.1 in
\cite{[Dob04]}. \tv{LIMQUA}=1 requires \tv{IPAHFB}=1.

\key{MULTLAGRAN} 0, 0, 0.0, 0 =
				\tv{LAMBDA}, \tv{MIU}, \tv{QLINEA}, \tv{IFLALQ}

\noindent  For \tv{IFLALQ}=1, the linear multipole constraint is used in conjunction
with the quadratic multipole constraint (see keyword \tv{MULTCONSTR}) to implement
the ALM for the total multipole moment constraint of multipolarity $\lambda$ and
$\mu$. The value of \tv{QLINEA}
is the initial value for the Lagrange parameter $L^{(0)}_{\lambda\mu}$. Updates of the
parameter in the ALM method are defined by Eq.~(\ref{eq:update}). The
calculated values of the Lagrange parameters are stored on the record
file; this allows for a smooth continuation of the ALM method when
restarting calculations from disk, see keyword \tv{CONTAUGMEN}.

For \tv{IFLALQ}=$-$1, only the linear multipole constraint is used for the multipolarity
$\lambda$ and $\mu$. This option is used together with \tv{IF\_RPA}=1. For
\tv{IFLALQ}=0, linear constraints are switched off.

\key{SURFLAGRAN} 0, 0, 0.0, 0 =
				\tv{LAMBDA}, \tv{MIU}, \tv{SLINEA}, \tv{IFLALS}

\noindent This keyword is the exact analog of \tv{MULTLAGRAN} for surface
and Schiff moments, see keywords \tv{SURFCONSTR} or \tv{SCHICONSTR})
in Sec.~2.4 of \cite{[Dob04]}. Additional values for the flag \tv{IFLALS} are possible:
for \tv{IFLALS}=2 or 3, the ALM is applied only for the neutron or proton
(surface-moment or Schiff-moment), respectively. The values of \tv{IFLALS}
must be the same for all constrained multipolarities.

\key{CONTAUGMEN} 0 =
				\tv{IACONT}

\noindent For \tv{IACONT}=1, the Lagrange parameters $L_{\lambda\mu}$ for the
linear constraints will be initialized with the values read from the record
file.

\key{RPA\_CONSTR} 0 =
				\tv{IF\_RPA}

\noindent For \tv{IF\_RPA}=1, the Lagrange parameters $L_{\lambda\mu}$ of the
linear constraints will be updated automatically at each iteration based on the
approximation of the RPA matrix. In \pr{hfodd} version (v\codeversion), this
option is only available for HFB calculations for conserved simplex (\tv{ISIMPY=1}). To be activated, it also requires
the flags for linear constraints {\tt IFLALQ} to be set (see
keyword \tv{MULTLAGRAN}). While in the ALM, these flags are all set to 1, in the
method based on the RPA matrix they must be set to $-$1.

\key{HFBTHOISON} 0, 0.0 =
				\tv{IF\_THO}, \tv{CBETHO}

\noindent For \tv{IF\_THO}=1, the code will automatically attempt to perform the
requested calculation, first with the \pr{hfbtho} solver, then by automatic
restart with the standard \pr{hfodd} engine. All options specific to \pr{hfodd}
and not implemented in \pr{hfbtho} will simply be disregarded in this first
stage. As an experimental feature, it is also possible to restart constrained
calculations, in which case \tv{CBETHO} is the value of the $\beta_{2}$
deformation used to start \pr{hfbtho}.

\key{BROYDENMAT} 4, 0 =
				\tv{NOIINP}, \tv{MIXMAT}

\noindent For \tv{MIXMAT}=1, the iterations of the self-consistent method
proceed by mixing the matrix elements of the HF(B) matrix instead of the HF
fields. This option is compatible with both \tv{IBROYD}=1 and \tv{IBROYD}=0. It
has been noticed that the mixing of matrix elements is less stable than the
mixing of the fields, unless the Broyden memory is erased every $n$ iterations.
The value of $n$ is given by \tv{NOIINP}, and it is recommended to take $n$ less
than the number of iterations kept in the Broyden memory.

\key{PARA\_ALL} 0, 1, 1, 1, 1 =
                       \tv{IPAALL}, \tv{NUBSTA}, \tv{NUBSTO}, \tv{NUTSTA}, \tv{NUTSTO}

\noindent For \tv{IPAALL}=1, calculations of kernels for different
values of the Euler angle $\beta$ and the gauge angle $\beta_T$
proceed in the same way as those for the $\alpha$ and $\gamma$ Euler
angles, see keyword \tk{PARAKERNEL} in Sec.~3.2 in \cite{[Dob09]}.
This allows for performing the calculation of kernels in parallel (in
different runs of the single-core version of \pr{hfodd}), and later using
the calculated kernels for the angular-momentum and isospin
projection. Calculations are performed for the nodes in the Euler
angle $\beta$ from {\tv{NUBSTA}} to {\tv{NUBSTO}} and for those in the
gauge angle $\beta_T$ from {\tv{NUTSTA}} to {\tv{NUTSTO}}. Values of
{\tv{NUBSTA}} and {\tv{NUBSTO}} must be between 1 and {\tv{NUBKNO}}
and must be ordered as {\tv{NUBSTA}}$\leq${\tv{NUBSTO}}. Values of
{\tv{NUTSTA}} and {\tv{NUTSTO}} must be between 1 and {\tv{NBTKNO}}
and must be ordered as {\tv{NUTSTA}}$\leq${\tv{NUTSTO}}.
\tv{IPAALL}=1 requires \tv{IPAKER}=1.

\key{NUMBKERNEL} 0 =
                       \tv{KFIKER}

\noindent For \tv{KFIKER}$>$0, the automated procedure of naming the
kernel files (see keyword \tk{SAVEKERNEL} in Sec.~3.2 in
\cite{[Dob09]}) is suspended and the kernels are saved in the kernel
file carrying the consecutive number equal to \tv{KFIKER}. This
requires an explicit bookkeeping of the kernel-file names in the input
data, but has the advantage of preventing two parallel jobs from
accessing the same kernel file simultaneously. Only the values of
\tv{KFIKER} between 0 and 999 are allowed. \tv{KFIKER}$>$0 requires
\tv{IPAKER}=1.


\subsection{High-Performance Computing}


\setcounter{mysubsubsection}{0}

\subsubsection{List of active keywords in hfodd.d}
\label{subsubsec:hfodd.d}

In parallel mode, the code \pr{hfodd} (v\codeversion) reads all user-defined
sequential data from the input file named \tv{hfodd.d}. Since this version is
the first to embed parallel capabilities, many \pr{hfodd} options have not been
implemented in parallel mode yet. Only a subset of \pr{hfodd} keywords can
therefore be activated, the list of which is given below:
\begin{itemize}
\item {\bf Iterations - } \tv{ITERATIONS}, \tv{BROYDEN}, \tv{SLOW\_DOWN},
\tv{SLOW\_PAIR}, \tv{SLOWLIPKIN}, \tv{ITERAT\_EPS}, \tv{MAXANTIOSC},
\tv{PING\_PONG}, \tv{CHAOTIC},
\item {\bf Specific features - }  \tv{FINITETEMP}, \tv{SHELLCORCT},
\tv{HFBTHOISON}, \tv{SHELLPARAM}, \tv{COULOMBPAR}, \tv{SKYRME-SET},
\tv{SKYRME\_STD}, \tv{UNEDF\_PROJ},
\item {\bf Constraints - } \tv{OMEGAY}, 
\item {\bf Symmetries - } \tv{SIMPLEXY}, \tv{SIGNATUREY}, \tv{PARITY},
\tv{ROTATION}, \tv{TSIMPLEX3D},
\item {\bf Pairing - } \tv{PAIRING}, \tv{HFB}, \tv{CUTOFF}, \tv{BCS},
\tv{HFBMEANFLD}, \tv{LIPKIN}, \tv{PAIR\_INTER}, \tv{PAIRNINTER},
\tv{PAIRPINTER},
\item {\bf HO Basis - } \tv{OPTI\_GAUSS}, \tv{GAUSHERMIT}, \tv{BASIS\_SIZE},
\tv{SURFAC\_DEF},
\item {\bf Miscellaneous - } \tv{ONE\_LINE}, \tv{NILSSONLAB}, \tv{REVIEW},
\item {\bf Restart options - } \tv{RESTART}, \tv{CONT\_PAIRI}, \tv{CONTLIPKIN},
\tv{CONTFIELDS}, \tv{EXECUTE}.
\end{itemize}

In principle, these options provide enough flexibility to cover the majority of
\pr{hfodd} applications in parallel mode. The user interested in some specific
option which could not be activated by one of the keywords above can still
manually modify the routine \tv{PREDEF} prior to compilation. This routine
pre-defines all \pr{hfodd} input data.


\subsubsection{Structure of hfodd\_mpiio.d}
\label{subsubsec:hfodd_mpiio.d}

\key{CALCULMODE} 1, 0 =
				\tv{MPIDEF}, \tv{MPIBAS}

\noindent For \tv{MPIDEF}=1, the code will perform a simple grid calculation of
$N_{\text{p}} \times N_{\text{n}} \times N_{\text{def}}$ points where
$N_{\text{p}}$ is the number of points along the $Z-$axis (proton number),
$N_{\text{n}}$ the number of points along the $N-$axis (neutron number), and
$N_{\text{def}}$ the total number of constraints on deformations, see keyword
\tv{MULTICONST} below. Requires \tv{IFCONS}=1. For \tv{MPIBAS}=1, the
calculation grid will be given by $ N_{\text{p}} \times N_{\text{n}} \times
N_{\text{def}} \times N_{\text{HO}} \times N_{\beta} \times N_{\omega} $, where
$N_{\text{HO}}$ is the number of different oscillator shells in the basis,
$N_{\beta}$ the number of different deformations of the basis, and
$N_{\omega}$ the number of different oscillator frequencies (in MeV), see also
keyword \tv{BASIS-NSHL}, \tv{BASIS-DEFS}  and \tv{BASIS-FREQ} below.

\key{CONSTR\_DEF} 1 =
				\tv{IFCONS}

\noindent For \tv{IFCONS}=1, every calculation performed by the code will be
constrained on the relevant values of the multipole moments.

\key{CONSTR\_LIN} 1 =
				\tv{IFLINE}

\noindent For \tv{IFLINE}=1, constrained calculations in multi-core mode 
are performed with the RPA method of re-adjusting the linear constraints, 
see Sec. \ref{sec:rpa}. 

\key{ALL\_NUCLEI} 66, 2, 1, 86, 2, 1 =
				\tv{IZSTRT}, \tv{IZSTEP}, \tv{NSTPEZ}, \tv{INSTRT}, \tv{INSTEP},
\tv{NSTEPN}

\noindent Define a vector of proton and neutron numbers
\begin{equation}
Z(i) = Z(0) + (i-1) \delta Z,\ \ i=1,\dots,N_{\text{p}}, \ \ \ \
N(j) = N(0) + (j-1)\delta N, \ \ j = 1,\dots, N_{\text{n}}.
\end{equation}
Then, $Z(0)$:=\tv{IZSTRT},  $\delta Z$:=\tv{IZSTEP},
$N_{\text{p}}$:=\tv{NSTEPZ},  $N(0)$:=\tv{INSTRT},  $\delta N$:=\tv{INSTEP},
$N_{\text{n}}$:=\tv{NSTEPN}. This defines a (rectangular) subset of nuclei in
the isotopic chart for which (possibly $N_{\text{def}}>1$) calculations will be
performed.

\key{MULTICONST} 2, 0, 10.0, 10.0, 4 =
				\tv{LAMBDA} \tv{MIU}, \tv{QBEGIN}, \tv{QFIN}, \tv{NUMBERQ}

\noindent The deformation grid is defined as a set of $N_{\text{def}} =
\Pi_{\lambda\mu} N_{\lambda\mu}$ deformation points where $N_{\lambda\mu}$ 
is the number of points for the constraint on the multipole moment
$\hat{Q}_{\lambda\mu}$ with multipolarity $\lambda$ (\tv{LAMBDA}) and $\mu$ 
(\tv{MIU}). The point number $k$ for this constraint reads:
\begin{equation}
\bar{Q}_{\lambda\mu}(k) = \bar{Q}_{\lambda\mu}(0) + \frac{k-1}{N_{\lambda\mu} - 1}
\left[ \bar{Q}_{\lambda\mu}(N_{\lambda\mu}) - \bar{Q}_{\lambda\mu}(0) \right],
\end{equation}
with $\bar{Q}_{\lambda\mu}(0)$:=\tv{QBEGIN}, 
$\bar{Q}_{\lambda\mu}(N_{\lambda\mu})$:=\tv{QFIN} and
$N_{\lambda\mu}$:=\tv{NUMBERQ}. Multiple constraints are obtained by adding
several lines with different $\lambda$ and $\mu$. All such lines must begin with
$\lambda < 0$ except the last one.

\key{BASIS-NSHL} 8, 2, 1 =
				\tv{N\_MINI}, \tv{N\_STEP}, \tv{NOFSHL}

\noindent For \tv{MPIBAS}=1, the number of shells in the basis
$N_{\text{HO}}$ can take different values of the form:
\begin{equation}
N_{\text{shell}}(m) = N_{\text{shell}}(0) + (m-1) \delta N_{\text{shell}}, \ \ m = 1,\dots,N_{\text{HO}}.
\end{equation}
Then, $N_{\text{shell}}(0) $:=\tv{N\_MINI},  $\delta N_{\text{shell}}$:=\tv{N\_STEP},
$N_{\text{HO}}$:=\tv{NOFSHL}. Note that the number of states is set independently
as a sequential data under keyword \tv{BASIS\_SIZE}. While the familiar relation
$N_{\text{states}} =
(N_{\text{shell}}+1)(N_{\text{shell}}+2)(N_{\text{shell}}+3)/6$ between the
number of states and the number of HO shells is valid for spherical bases, it
does not apply in the deformed case.

\key{BASIS-DEFS} 0.0, 0.1, 1 =
				\tv{B20MIN}, \tv{B20STP}, \tv{NOFB20}

\noindent For \tv{MPIBAS}=1, the axial quadrupole deformation $\beta\equiv
\alpha_{20}$ can take different values of the form:
\begin{equation}
\beta(n) = \beta(0) + (n-1)\delta\beta, \ \ n = 1,\dots,N_{\beta}.
\end{equation}
Then, $\beta(0)$:=\tv{B20MIN},  $\delta \beta$:=\tv{B20STP},
$N_{\beta}$:=\tv{NOFB20}

\key{BASIS-FREQ} 8.0, 0.1, 1 =
				\tv{O\_MINI}, \tv{O\_STEP}, \tv{NOFFRE}

\noindent For \tv{MPIBAS}=1, the oscillator frequency $\hbar\omega$ can take 
different values of the form:
\begin{equation}
\hbar\omega(l) = \hbar\omega(0) + (l-1)\delta\omega, \ \ l=1,\dots,N_{\omega}.
\end{equation}
Then, $\hbar\omega(0)$:=\tv{O\_MINI},  $\delta \omega$:=\tv{O\_STEP},
$N_{\omega}$:=\tv{NOFFRE}


\section{Output Files}
\label{sec:output}

Four additional examples of output file, illustrating the new features of the
code are provided in files \tv{ca40\_iso.out}, \tv{cr48\_lip.out}, \tv{cr50\_tem.out} 
and \tv{pb208\_tho.out}. Selected lines from \tv{ca40\_iso.out} are given in 
Appendix B below. Several minor sections of the output have been added or 
reformatted. We describe below only the non-trivial important additions.

In \tv{ca40\_iso.out}, the section labeled \tv{ISOSPIN-MIXED EIGENSTATES} 
lists all the eigenvectors (within the $\epsilon_{T}$ cut-off described in
 Sec. \ref{subsubsec:isospin}) of the Hamiltonian re-diagonalized in the 
 good-isospin basis. The first two columns are the number $n$ and value 
 $E_{n,T_{z}}$ of the energy. The next 5 columns give the characteristics of the 
 expansion of Eq.(\ref{mix2a}) (or Eq.(\ref{KTmix}) if isospin and angular 
 momentum are combined). The fourth and fifth columns give respectively the 
 number $m$ and isospin $T$ of the good-isospin basis state. The columns 
 6, 7 and 8 give, respectively, the norm of the expansion coefficient 
 $a_{mT}^{(n)}$, its real part and and its imaginary part. 

In \tv{cr50\_tem.out}, the new value \tv{I\_LINE=3} has been used to display the 
iterations. Every line is made of the iteration number \tv{ITER}, the value of the 
total energy \tv{ENERGY}, the value of the stability criterion \tv{STABILITY}, the 
total quadrupole moment \tv{Q\_2} the $\gamma$ angle \tv{GAMMA}, the total 
entropy \tv{S}, the neutron Fermi level \tv{lam\_n}, the proton Fermi level 
\tv{lam\_p}, the neutron pairing energy \tv{EpN} and the proton pairing energy 
\tv{EpP}. With the values of the Fermi levels, total entropy and temperature, the 
grand canonical potential $\Omega$ of Eq.(\ref{grpot}) can be deduced at each iteration.

In \tv{pb208\_tho.out}, a warning message is displayed at the beginning to 
indicate that the calculation will proceed in two steps, first with \pr{hfbtho} 
then with \pr{hfodd}. The message also gives the conditions under which 
the restart is expected to be smooth. Follows the output generated by 
\pr{hfbtho}, we refer to \cite{[Sto05]} for additional information. Since the 
UNEDF functional is used in this test run, an additional section \tv{NUCLEAR 
MATTER PROPERTIES} lists the volume nuclear matter characteristics of the 
functional. Section \tv{THE SHELL CORRECTION...} gives the type of shell 
correction computed and the numerical characteristics of the smoothing 
procedure. Section \tv{SHELL CORRECTION} located just before the final 
energy table gives the value of the shell correction for neutrons and protons.


\section{Fortran Source Files}
\label{sec:source}

The Fortran source code is provided as a set of 8 module files:
\begin{itemize}
\item \tv{hf249t.f} - Main source code
\item \tv{hfodd\_functional.f90} - Definition of energy functionals based on
nuclear matter properties and coupling constants instead of (t,x) parameters.
\item \tv{hfodd\_shell.f} - Shell correction
\item \tv{hfodd\_hfbtho.f90} - \pr{hfbtho} code (v.101a)
\item \tv{hfodd\_interface.f90} - Interface between \pr{hfbtho} and \pr{hfodd}
\item \tv{hfodd\_mpiio.f90} - Module handling I/O in parallel mode
\item \tv{hfodd\_mpimanager.f90} - Module defining parallel jobs
\item \tv{hfodd\_SLsiz.f90} - Scalapack module for routine \tv{HFBSIZ}
\end{itemize}
together with one header file, \tv{hfodd\_sizes.h}, which contains all Fortran
\tv{PARAMETER} statements controlling the size of static arrays. The language of
newly-developed modules is Fortran 90, while legacy code is still written,
in part or totally, in Fortran 77.

\subsection{Standard Libraries}

The code \pr{hfodd} requires an implementation of the BLAS and LAPACK
libraries to function correctly, see Sec.~5.2 in \cite{[Dob09]} for details. While
the interface to older NAGLIB routines remains available, BLAS and LAPACK
will give the best peak performance on most current computers and are
recommended.

\setcounter{mysubsubsection}{0}

\subsection{Parallel Mode}

We recall that a parallel machine is made of a certain number of sockets, each
containing one processor. Every processor contains a number of CPU units, or
cores, sharing the same memory.

\subsubsection{Basic MPI}

To activate multi-core calculations, \pr{hfodd} requires an implementation of
the Message Passing Interface (MPI). The current version was tested on two
different implementations:
\begin{itemize}
\item MPICH-1 and MPICH-2, available at:

http://www.mcs.anl.gov/research/projects/mpich2/
\item Open MPI available at: http://www.open-mpi.org/
\end{itemize}
In parallel mode, the code \pr{hfodd} is compiled by setting \tv{USE\_MPI} to
1 in the project Makefile. Typically, the executable is run as follows (\tv{bash} syntax):\\

\tv{mpiexec -np [number of processes] hf249t < /dev/null >\& hf249t.out}\\

\noindent where \tv{hf249t.out} is a redirection for the standard output and
files \tv{hfodd.d} and \tv{hfodd\_mpiio.d} must be in the directory where this
command is run.

\subsubsection{Hybrid OpenMP/MPI Mode}

Multi-threading is activated by switching the \tv{USE\_OPENMP} to 1 in the project
Makefile. This option can be used on its own, or in combination with \tv{USE\_MPI}=1,
in which case the programming model is hybrid MPI/OpenMP. We recall that
to activate multi-threading, the environment variable \tv{OMP\_NUM\_THREADS}
must be set to the required number of threads prior to execution. If every processor
has 6 cores, then to run 12 MPI processes with 3 threads each, the following command
line (in the OPENMPI implementation) should be executed:\\

\tv{export OMP\_NUM\_THREADS = 3}\\

\tv{mpiexec -np 12 -npersocket 2 hf249t < /dev/null >\& hf249t.out}\\

\noindent Therefore, instead of the 12 MPI processes being executed by all the 12 cores
of 2 full processors, the \tv{-npersocket 2} option imposes that only 2 cores within
a given socket are actually used, leaving the remaining 4 available when multi-threading
kicks in. Such an instruction requires 6 processors instead of 2 in the pure MPI
mode, and up to (12 processes)$\times$ (3 threads) = 36 cores may be active at a 
given time .

\subsubsection{Scalapack}

Using Scalapack requires the most advanced partitioning of the core grid. The library
can be downloaded at:\\

http://www.netlib.org/scalapack/\\

\noindent It relies on the BLACS framework, which is available at \\

http://www.netlib.org/blacs\\

\noindent To compile the code with the Scalapack library, switch
\tv{USE\_SCALAPACK} to 1 in the project Makefile. Note that Scalapack
requires a multi-core grid and can therefore not be used in serial mode:
it requires to set \tv{USE\_MPI} to 1 in the Makefile. Optimal
performance can be obtained by also allowing multi-threading. The
syntax of the command line is unchanged compared to the basic MPI
or hybrid model. However, special care must be taken in the choice of
the number of cores: the total number of cores is now the product of
the number of cores for each HFB calculation (size of the Scalapack
process grid) by the number of different HFB calculations required
(defined in \tv{hfodd\_mpiio.d}).


\section{Acknowledgments}
\label{sec:acknowledgments}

\bigskip
We thank Micha{\l} Opala for performing OpenMP tests and bringing our
attention to Amdahl's law. Discussions with Hai Ah Nam on scaling properties
on leadership class computers are also warmly acknowledged. This work was
supported in part by the Polish Ministry of Science and Higher Education under
Contract Nos.~N~N202~328234 and N~N202~231137, by the Academy of Finland and University of
Jyv\"askyl\"a within the FIDIPRO program, by the UNEDF SciDAC Collaboration
under the U.S.\ Department of Energy grants No.~DE-FC02-07ER41457 and
DE-FG02-96ER40963 (University of Tennessee), and was partly performed
under the auspices of the US Department of Energy by the Lawrence Livermore
National Laboratory under Contract DE-AC52-07NA27344 
(code release number: LLNL-CODE-470611, 
d ../.document release number: LLNL-JRNL-472093). 
Funding was also
provided by the United States Department of Energy Office of Science, Nuclear
Physics Program pursuant to Contract DE-AC52-07NA27344 Clause B-9999,
Clause H-9999 and the American Recovery and Reinvestment Act, Pub. L. 111-5.
Computational resources were provided in part by a computational grant from
the Interdisciplinary Centre for Mathematical and Computational Modeling (ICM)
of the Warsaw University, by the Oak Ridge Leadership Computing Facility,
located in the National Center for Computational Sciences at Oak Ridge
National Laboratory supported by the Office of Science of the U.S. Department
of Energy under Contract DE-AC05-00OR22725, as well as by the National
Energy Research Scientific Computing Center supported by the Office of
Science of the U.S. Department of Energy under Contract
No. DE-AC02-05CH11231.
We also acknowledge the CSC - IT Center for Science Ltd, Finland for the
allocation of computational resources.


\appendix

\section{Test Run Input}

\begin{verbatim}
!-----------------------------------------------------------------------------!
! This file is part of the official HFODD v2.49t release and demonstrates     !
! the use of isospin mixing and projection.                                   !
!-----------------------------------------------------------------------------!

                      ----------  General data  -----------
NUCLIDE      IN_FIX  IZ_FIX
               20      20
ITERATIONS   NOITER
              100     
ITERAT_EPS   EPSITE
            0.000000001
SLOW_DOWN    SLOWEV  SLOWOD
               0.5     0.5     
PRINT_ITER   IPRSTA  IPRMID  IPRSTO
                0        0     1
MAXANTIOSC   NULAST
                3
BROYDEN      IBROYD  N_ITER  ALPHAM  BROTRI
                1      7       0.3  10000.0
                
                      -----------  Interaction  -----------
SKYRME-SET   SKYRME
            SV  
SKYRME-STD   ISTAND  KETA_J  KETA_W  KETACM  KETA_M
               0        1       0       0       1
LANDAU       LANODD  X0_LAN  X1_LAN  G0_LAN  G0PLAN  G1_LAN  G1PLAN
              000      1.0     1.0     0.4     1.2    -0.19   0.62
EVE_SCA_TS   RHO       RHOD      LPR       TAU       SCU       DIV
            1.  1.    1.  1.    1.  1.    1.  1.    1.  1.    1.  1.
ODD_SCA_TS   SPI       SPID      LPS       CUR       KIS       ROT
            1.  1.    1.  1.    1.  1.    1.  1.    1.  1.    1.  1.
EVE_SCA_PM   RHO       RHOD      LPR       TAU       SCU       DIV
            1.  1.    1.  1.    1.  1.    1.  1.    1.  1.    1.  1.
ODD_SCA_PM   SPI       SPID      LPS       CUR       KIS       ROT
            1.  1.    1.  1.    1.  1.    1.  1.    1.  1.    1.  1.
            
                      ----------    Pairing    ------------
PAIRING      IPAIRI
               0
HFB          IPAHFB
               0

                     -----------  Symmetries  -------------
ROTATION     IROTAT
               1
SIMPLEXY     ISIMPY
               0
SIGNATUREY   ISIGNY
               0
PARITY       IPARTY
               0
TSIMPLEX3D   ISIMTX  ISIMTY  ISIMTZ
               0       0       0
               
                     ----------  Configurations  ----------
PHNONE_NEU            PARTICLES                 HOLES
            1            000                     000
PHNONE_PRO            PARTICLES                 HOLES
            1            000                     000
DIANON_NEU
            24  23  0        
DIANON_PRO
            24  23  0        
VACSIM_NEU            SIMP SIMM
                       12   12
VACSIM_PRO            SIMP SIMM
                       10   10
VACPAR_NEU            PARP PARM
                       14   10
VACPAR_PRO            PARP PARM
                       14    6
                  
                     ----  Parameters of the HO basis  ----
BASIS_SIZE   NOSCIL  NLIMIT  ENECUT
               10     286    800.0
SURFAC_PAR   INNUMB  IZNUMB  R0PARM
               20      20     1.23
OPTI-GAUSS   IOPTGS
                1
GAUSHERMIT   NXHERM  NYHERM  NZHERM
               26      26      26
SURFAC_DEF   LAMBDA   MIU    ALPHAR
               -2      0      0.0
                4      0      0.0
               
                     -----------  Constraints  ------------
OMEGAY       OMEGAY
              0.00 
OMEGA_XYZ    OMEHAX  OMEHAY  OMEHAZ  ITILAX
              0.000   0.000   0.000    0
MULTCONSTR   LAMBDA    MIU   STIFFQ  QASKED  IFLAGQ
               -2       0     0.25    0.200     1
                2       2     0.25    0.000     1
                
                     ------  Output-file  parameters  -----
EALLMINMAX   EMINAL  EMAXAL
              -30.0   10.0
              
                     --------------  Files  ---------------
REVIEWFILE   FILREV
            ca40_iso.rev
RECORDFILE   FILREC
            ca40_iso.rec                                                   
REPLAYFILE   FILREP
            ca40_iso.rec                                      
REC_FIELDS   FILFIC
            ca40_iso.fil
REP_FIELDS   FILFIP
            ca40_iso.fil
COULOMFILE   FILCOU
            ca40_iso.cou
REVIEW       IREVIE
                0
RECORDSAVE   IWRIRE
                1      
COULOMSAVE   ICOULI  ICOULO
                1       1
FIELD_SAVE   IWRIFI
                1
FIELD_OLD    IWRIOL
                1
                     ------  Starting the iteration  ------
RESTART      ICONTI
               0 
CONTFIELDS   IFCONT
               0 
CONT_PAIRI   IPCONT
               0
               
                     ------------  Calculate  -------------
EXECUTE

                     ------------  Next run   -------------
ITERATIONS   NOITER
              30     
SLOW_DOWN    SLOWEV  SLOWOD
               0.5     0.5     
BROYDEN      IBROYD  N_DUMM  ALPHAM  BROTRI
                0      7       0.3  10000.0              
MULTCONSTR   LAMBDA    MIU   STIFFQ  QASKED  IFLAGQ
               -2       0     0.25    0.200     0
                2       2     0.25    0.000     0                
COULOMBPAR   ICOTYP  ICOUDI  ICOUEX
               5        2      2
RESTART      ICONTI
               1 
CONTFIELDS   IFCONT
               1 
CONT_PAIRI   IPCONT
               0
               
                     ------------  Calculate  -------------
EXECUTE

                     ------------  Next run   -------------
ITERATIONS   NOITER
               1     
SLOW_DOWN    SLOWEV  SLOWOD
               1.0     1.0               
KERNELFILE   FILKER
            ca40_iso.ker
SAVEKERNEL   ISAKER
               1
PARAKERNEL   IPAKER  NUASTA  NUASTO  NUGSTA  NUGSTO
               0       1       1       1       1
PROJECTGCM   IPRGCM  IPROMI  IPROMA  NUAKNO  NUBKNO  KPROJE  IFRWAV  ITOWAV  IWRWAV
               1       0       0       1       1       0       1       1       0
CUTOVERLAP   ICUTOV  CUTOVE  CUTOVF
               1     1.0E-05   1.0
PROJECTISO   IPRGCM  ISOSAD  NBTKNO  EPSISO  ICSKIP  IFERME
               1       10      8     1.D-10     0       0               
RESTART      ICONTI
               1 
CONTFIELDS   IFCONT
               1 
CONT_PAIRI   IPCONT
               0
               
                     ------------  Calculate  -------------
EXECUTE
                     ------------  Terminate  -------------
ALL_DONE
\end{verbatim}

\section{Test Run Output}

\begin{verbatim}
*******************************************************************************
*                                                                             *
*                S I N G L E - C O R E    V E R S I O N                       *
*                                                                             *
*******************************************************************************
*                                                                             *
*     HFODD    HFODD    HFODD    HFODD    HFODD    HFODD    HFODD    HFODD    *
*                                                                             *
*******************************************************************************
*                                                                             *
*             SKYRME-HARTREE-FOCK-BOGOLYUBOV CODE VERSION: 2.49T              *
*                                                                             *
*              NO SYMMETRY-PLANES AND NO TIME-REVERSAL SYMMETRY               *
*                                                                             *
*                DEFORMED CARTESIAN HARMONIC-OSCILLATOR BASIS                 *
*                                                                             *
*******************************************************************************
*                                                                             *
*              J. DOBACZEWSKI, B.G. CARLSSON, J. DUDEK, J. ENGEL              *
*           J. MCDONNELL, P. OLBRATOWSKI, P. POWALOWSKI, M. SADZIAK           *
*                J. SARICH, W. SATULA, N. SCHUNCK, J.A. SHEIKH                *
*             A. STASZCZAK, M. STOITSOV, P. TOIVANEN, M. ZALEWSKI             *
*                               AND H. ZDUNCZUK                               *
*                                                                             *
*                   INSTYTUT FIZYKI TEORETYCZNEJ, WARSZAWA                    *
*                 LAWRENCE LIVERMORE NATIONAL LABORATORY, USA                 *
*                                                                             *
*                                  1993-2011                                  *
*                                                                             *
*******************************************************************************

*******************************************************************************
*                                                                             *
*  CODE COMPILED WITH THE FOLLOWING ARRAY DIMENSIONS AND SWITCHES:            *
*                                                                             *
*******************************************************************************
*                                                                             *
*  NDBASE =  680  NDSTAT =  680  NDXHRM =   40  NDYHRM =   40  NDZHRM =   40  *
*                                                                             *
*  NDMAIN =   16  NDMULT =    9  NDMULR =    4  NDLAMB =    9  NDITER = 5000  *
*                                                                             *
*  NDAKNO =    1  NDBKNO =    1  NDPROI =   20  NDCOUL =   80  NDPOLS =   25  *
*                                                                             *
*  NDPROT =   10  NDBTKN =   10                                               *
*                                                                             *
*  IPARAL =    0  I_CRAY =    0                                               *
*                                                                             *
*                                                                             *
*  PRE-PROCESSOR OPTIONS:                                                     *
*                                                                             *
*       switch_port = 1     switch_diag = 3     switch_cray = 0               *
*                                                                             *
*       switch_nagl = 0     switch_quad = 0     switch_vect = 1               *
*                                                                             *
*******************************************************************************
\end{verbatim}

\centering $\vdots$

\begin{verbatim}
*******************************************************************************
*                                                                             *
* BROYDEN METHOD IS:  ON                                                      *
*                                                                             *
*    TRIGGERED ONLY WHEN STABILITY IS LOWER THAN : 10000.000 MEV              *
*    INITIAL SLOWING FACTOR (BEFORE TRIGGER)     :     0.50 (=SLOWEV)         *
*    BROYDEN SLOWING FACTOR (AFTER  TRIGGER)     :     0.70 (=1-ALPHAM)       *
*    NUMBER OF ITERATIONS RETAINED IN MEMORY     :     7                      *
*                                                                             *
*******************************************************************************
\end{verbatim}

\centering $\vdots$

\begin{verbatim}
*******************************************************************************
*                                                                             *
*                  ONLY THE ISOSPIN PROJECTION IS PERFORMED                   *
*                        8 GAUSS-LEGENDRE KNOTS IS USED                       *
*                                                                             *
*******************************************************************************
\end{verbatim}

\centering $\vdots$

\begin{verbatim}
*******************************************************************************
*                                                                             *
*      CUT-OFF "EPSISO" =  0.00000000010000 ==> GOOD-T BASIS OF DIM =  4      *
*                                                                             *
*******************************************************************************
*                                                                             *
*                          ISOSPIN-MIXED EIGENSTATES                          *
*                          -------------------------                          *
*                                                                             *
*     N       EIGENENERGY     i    T       |C_i|^2     Re[C_i]     Im[C_i]    *
*                                                                             *
*     1       -342.860677     1    0      0.994703    0.997348    0.000000    *
*                             2    1      0.005296    0.072772    0.000000    *
*                             3    2      0.000002    0.001316    0.000000    *
*                             4    3      0.000000   -0.000061    0.000000    *
* DOMINANT AMPLITUDE SQUARED EQUALS:  0.9947025220 AND CORRESPONDS TO T = 0   *
* COULOMB MIXING IN  THIS  STATE IS:  0.0052974780 [  0.529748%]              *
*                                                                             *
*                                                                             *
*     2       -300.253660     1    0      0.005276    0.072636    0.000000    *
*                             2    1      0.988062   -0.994013    0.000000    *
*                             3    2      0.006659   -0.081605    0.000000    *
*                             4    3      0.000003   -0.001655    0.000000    *
* DOMINANT AMPLITUDE SQUARED EQUALS:  0.9880618341 AND CORRESPONDS TO T = 1   *
* COULOMB MIXING IN  THIS  STATE IS:  0.0119381659 [  1.193817%]              *
*                                                                             *
*                                                                             *
*     3       -258.016938     1    0      0.000021   -0.004628    0.000000    *
*                             2    1      0.006609    0.081296    0.000000    *
*                             3    2      0.985082   -0.992513    0.000000    *
*                             4    3      0.008288   -0.091039    0.000000    *
* DOMINANT AMPLITUDE SQUARED EQUALS:  0.9850815887 AND CORRESPONDS TO T = 2   *
* COULOMB MIXING IN  THIS  STATE IS:  0.0149184113 [  1.491841%]              *
*                                                                             *
*                                                                             *
*     4       -215.384252     1    0      0.000000    0.000241    0.000000    *
*                             2    1      0.000033   -0.005784    0.000000    *
*                             3    2      0.008257    0.090869    0.000000    *
*                             4    3      0.991709   -0.995846    0.000000    *
* DOMINANT AMPLITUDE SQUARED EQUALS:  0.9917092280 AND CORRESPONDS TO T = 3   *
* COULOMB MIXING IN  THIS  STATE IS:  0.0082907720 [  0.829077%]              *
*                                                                             *
*                                                                             *
*******************************************************************************
\end{verbatim}


\bibliography{hfodd}
\bibliographystyle{cpc}

\end{document}